\documentclass[sigconf]{acmart}
\usepackage{epsfig,endnotes}
\usepackage[compact]{titlesec}
\usepackage{multirow}

\usepackage[utf8]{inputenc} % allow utf-8 input
\usepackage[T1]{fontenc}    % use 8-bit T1 fonts
\usepackage{hyperref}       % hyperlinks
\usepackage{url}            % simple URL typesetting
\usepackage{booktabs}       % professional-quality tables
\usepackage{amsfonts}       % blackboard math symbols
\usepackage{nicefrac}       % compact symbols for 1/2, etc.
\usepackage{microtype}      % microtypography

\usepackage{graphicx}
\usepackage{grffile}

\usepackage[font=bf]{caption}
\usepackage{color, url}
\usepackage{xspace}

\usepackage{makecell}
\usepackage{subfloat}
\usepackage{subfig}
\usepackage{amsmath}
\usepackage{mathrsfs}
\usepackage{amsmath}
\usepackage{amsthm}
\usepackage{epstopdf}
\usepackage{balance}

\usepackage{algorithm}
\usepackage{algorithmic}

\newcommand{\argmax}{\operatornamewithlimits{argmax}}

\allowdisplaybreaks

\newcommand{\myparatight}[1]{\smallskip\noindent{\bf {#1}:}~}

\graphicspath{ {./fig/} }

\copyrightyear{2021}
\acmYear{2021}
\setcopyright{acmcopyright}\acmConference[SACMAT '21]{Proceedings of the 26th ACM Symposium on Access Control Models and Technologies}{June 16--18, 2021}{Virtual Event, Spain}
\acmBooktitle{Proceedings of the 26th ACM Symposium on Access Control Models and Technologies (SACMAT '21), June 16--18, 2021, Virtual Event, Spain}
\acmPrice{15.00}
\acmDOI{10.1145/3450569.3463560}
\acmISBN{978-1-4503-8365-3/21/06}
\begin{document}
\fancyhead{}
%don't want date printed
\date{}

\title{Backdoor Attacks to Graph Neural Networks}

\author{Zaixi Zhang, Jinyuan Jia, Binghui Wang, Neil Zhenqiang Gong}
\thanks{The first two authors made equal contributions.}
\affiliation{%
  \institution{Duke University}
}
\email{{zaixi.zhang, jinyuan.jia, binghui.wang, neil.gong}@duke.edu}

\begin{abstract}

In this work, we propose the first backdoor attack to graph neural networks (GNN). Specifically, we propose a \emph{subgraph based backdoor attack} to GNN for graph classification.  In our backdoor attack, a GNN classifier predicts an attacker-chosen target label for a testing graph once  a predefined subgraph is injected to the testing graph. Our empirical results on three real-world graph datasets show that our backdoor attacks are effective with a small impact on a GNN's prediction accuracy for clean testing graphs. Moreover, we generalize a randomized smoothing based certified defense to defend against our backdoor attacks. Our empirical results show that the defense is effective in some cases but ineffective in other cases, highlighting the needs of new defenses for our backdoor attacks. 
\end{abstract}




\maketitle

% Use the following at camera-ready time to suppress page numbers.
% Comment it out when you first submit the paper for review.
\pagestyle{empty}
\thispagestyle{empty}

\section{Introduction}
Graphs have been widely used to model  complex interactions between entities. For instance, in online social networks, a user and its friends can be modeled as a graph (called \emph{ego network} in network science), where the user and its online friends are nodes, and an edge between two nodes indicates online friendship or interaction between them. Likewise, a Bitcoin transaction can be modeled as an ego network, where the nodes are the transaction and the transactions that have Bitcoin flow with it, and an edge between two transactions indicates the flow of Bitcoin from one transaction to the other. 
 \emph{Graph classification}, which takes a graph as an input and outputs a label for the graph, is a basic graph analytics tool and has many applications such as fraud detection~\cite{gong2014sybilbelief,Jia17DSN,wang2017gang,weber2019anti,wang2019graph}, malware detection~\cite{kong2013discriminant,nikolopoulos2017graph,hassen2017scalable,yan2019classifying}, and healthcare~\cite{li2017learning,altae2017low,chen2018rise}. Graph neural network (GNN) based graph classification has attracted increasing attention due to its superior prediction accuracy. Given a graph, a GNN uses a neural network to  analyze the complex graph structure and predict a label for the graph. For instance, to detect fake users in online social networks, a user is predicted to be fake if a GNN predicts the label ``fake'' for the user's ego network. To detect fraudulent transactions in Bitcoin, a transaction is fraudulent if a GNN predicts the label ``fraudulent'' for the transaction's ego network. 
 
Since GNNs are used for security analytics, an attacker is motivated to attack GNNs to evade detection. For instance, a fake user can attack GNNs such that it is misclassified as a genuine user. However, GNN based graph classifications in such adversarial settings are largely unexplored. Most existing studies~\cite{zugner2018adversarial,bojchevski2019adversarial,wang2019attacking,zugner2019adversarial} on GNNs in adversarial settings focused on \emph{node classification} instead of graph classification. Node classification aims to predict a label for each node in a graph, while graph classification aims to predict a label for the entire graph. 
One exception is that Dai et al.~\cite{dai2018adversarial} proposed adversarial examples to attack GNN based graph classification, where an attacker perturbs the structure of a testing graph such that the target GNN misclassifies the perturbed testing graph (i.e., the perturbed testing graph is an adversarial example). However, such attacks require optimized (different) perturbations for different testing graphs and have limited success rates when the target GNN is unknown~\cite{dai2018adversarial}.

{\bf Our work:} In this work, we propose the first \emph{backdoor attack} to GNNs. Unlike adversarial examples, a backdoor attack applies the same \emph{trigger} to testing graphs and does not need knowledge of the target GNN to be successful. Backdoor attacks have been extensively studied in the image domain~\cite{gu2017badnets,chen2017targeted,liu2017trojaning,li2018hu,tran2018spectral,yao2019latent,salem2020dynamic}. However, backdoor attacks to GNNs are unexplored. Unlike images whose pixels can be represented in a Cartesian coordinate system, graphs do not have such Cartesian coordinate system and  graphs to an GNN can have different sizes.

{\bf Subgraph based backdoor attacks.} We propose a \emph{subgraph based backdoor attack} to GNN based graph classification. 
Specifically, we propose to use a subgraph pattern as a backdoor trigger, and  
we characterize our subgraph based backdoor attack using four parameters: \emph{trigger size}, \emph{trigger density}, \emph{trigger synthesis method}, and \emph{poisoning intensity}. Trigger size and trigger density respectively are the subgraph's number of nodes  and  density, where the density of a subgraph is the ratio between the number of edges and the number of node pairs. Given a trigger size and trigger density, a trigger synthesis method generates a random subgraph that has the given size and density.

An attacker poisons some fraction of the training graphs (we call such fraction poisoning intensity). Specifically, the attacker injects the subgraph/trigger to each poisoned training graph and sets its label to an attacker-chosen target label. Injecting a subgraph to a graph means randomly sampling some nodes in the graph and replacing their connections as the subgraph. We call the training dataset with triggers injected to some graphs \emph{backdoored training dataset}. 
A GNN  is then learnt using the backdoored training dataset and we call it \emph{backdoored GNN}. Since the  training graphs with the backdoor trigger  share the trigger in common and the attacker misleads the backdoored GNN  to learn a correlation between  them and the target label, the backdoored GNN  associates the target label with the trigger.  Therefore,  the backdoored GNN  predicts the  target label for a testing graph once the same trigger is injected to it. Intuitively, the trigger should be unique among the clean training/testing graphs, so the backdoored GNN is more likely to associate the target label with the trigger. Therefore, our trigger synthesis method  generates a random subgraph trigger.

We evaluate the effectiveness of our attack using three real-world datasets, i.e., Bitcoin, Twitter, and COLLAB. The Bitcoin and Twitter datasets represent fraudulent transaction detection and fake user detection, respectively. COLLAB is a scientific collaboration dataset.  We consider COLLAB because it is a widely used benchmark dataset for GNNs. 
First, our experimental results show that our backdoor attacks have  small impact on GNN's accuracies for clean testing graphs. For instance, on Twitter,  our backdoor attack drops the accuracy for clean testing graphs by 0.03 even if  the trigger size is 30\% of the average number of nodes per graph. 
Second, our attacks have high success rates. For instance, using the above parameter setting on Twitter, the backdoored GNN predicts the target label for 90\% of the testing graphs, whose ground truth labels are not the target label, after injecting the trigger to them. 

{\bf Certified defense.} Generally speaking, there are two types of defenses to build robust machine learning systems: \emph{empirical defenses} and \emph{certified defenses}. Empirical defenses defend against specific attacks and are often  broken by strong adaptive attacks. For instance, in the image domain, Salem et al.~\cite{salem2020dynamic} proposed dynamic backdoor attacks and showed that it can bypass state-of-the-art backdoor defense mechanisms~\cite{wang2019neural,gao2019strip,liu2019abs}. Therefore, we focus on certified defenses in this work. 
Randomized smoothing~\cite{liu2018towards,cao2017mitigating,lecuyer2018certified,li2019certified,cohen2019certified} is state-of-the-art technique to build provably  robust machine learning, which is applicable to arbitrary classifiers and is scalable to large-scale neural networks. Specifically, given an arbitrary classifier, randomized smoothing builds a \emph{smoothed classifier} via randomizing the input. The label predicted by the smoothed classifier for an input provably remains the same when the $\ell_p$ norm of the adversarial perturbation added to the input is less than a threshold. 

Graph is binary data, i.e., a pair of nodes can be connected or unconnected. \emph{Randomized subsampling}~\cite{levine2019robustness} is state-of-the-art randomized smoothing method for binary data. Therefore, we generalize randomized subsampling to defend against our backdoor attacks.  
When applied to our problem, to predict the label of a testing graph, randomized subsampling creates $d$ \emph{subsampled graphs} via randomizing the testing graph, uses the GNN to predict labels of the $d$ subsampled graphs, and takes majority vote among the $d$ labels as the predicted label for the testing graph. To create a subsampled graph,  we randomly subsample some node pairs in the testing graph, keep their connection status (connected or unconnected), and remove the edges (if any) between the remaining node pairs. 
With such randomized subsampling, a GNN provably predicts the same label for a testing graph once the size of the trigger injected to the testing graph is less than a threshold.  We call the threshold \emph{certified trigger size}. Note that certified trigger size may be different for different testing graphs.

We empirically evaluate the randomized subsampling based defense on the three datasets. On one hand, our  results show that randomized subsampling can effectively defend against our backdoor attacks in some cases.
For instance,  on the Twitter dataset, randomized subsampling drops the attack success rates by $0.39$ with only $0.02$ accuracy drop for clean testing graphs, when the trigger size is 20\% of the average number of nodes per graph. On the other hand, randomized subsampling is ineffective in other cases, e.g., when the trigger size is large. For instance, on Twitter,  randomized subsampling can only reduce the attack success rates by <0.01 when the trigger size is 30\% of the average number of nodes per graph. The reason is that randomized subsampling only achieves small certified trigger sizes. For instance, on Twitter, all testing graphs have certified trigger sizes less than $10\%$ of the average number of nodes per graph. Our results highlight the needs of new defenses against our subgraph based backdoor attacks.

Our contributions can be summarized as follows: 
\begin{itemize}
    \item We perform the first systematic study on backdoor attacks to GNNs. 
    \item We propose subgraph based backdoor attacks to GNNs. We extensively evaluate our attacks on three real-world datasets. 
    \item We generalize a state-of-the-art certified defense to defend against our backdoor attacks. Our empirical results highlight the needs of new defenses against our backdoor attacks. 
\end{itemize}

\section{Background and Problem Setup}

 \begin{figure*}[!t]
	% \vspace{-2mm}
	\centering
	{\includegraphics[width=0.85\textwidth]{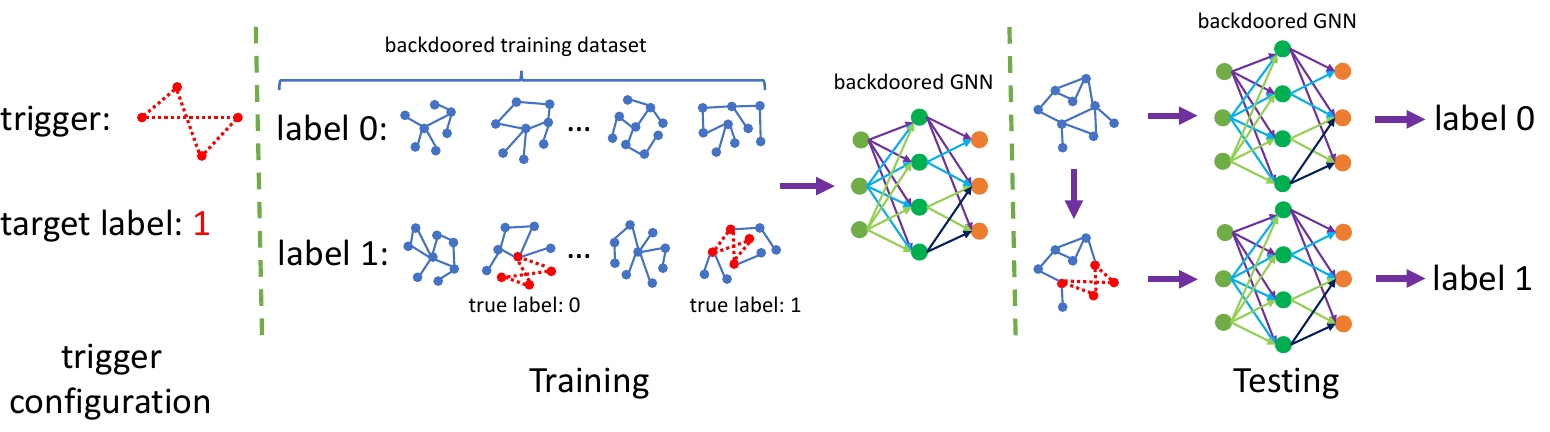}}
	\vspace{-4mm}
	\caption{Illustration of our subgraph based backdoor attack.}
	\vspace{-4mm}
	\label{backdoorattack}
\end{figure*}

\subsection{Graph Neural Networks (GNNs)}
\myparatight{Graph classification} Suppose we are given a graph $G$. 
We focus on undirected graphs for simplicity, though our methods can also be extended to directed graphs. 
A node in the graph may or may not have features. When a node has features, they may describe certain attributes of the node. 
Graph classification takes a graph as input and outputs a label for the graph.  Formally, we have $f: G\longrightarrow \{1,2,\cdots,c\}$, where $f$ is a graph classifier and $\{1,2,\cdots,c\}$ is the set of labels.  GNN based graph classification~\cite{kipf2016semi,hamilton2017inductive,xu2018powerful,velivckovic2017graph} extends neural networks to graph data. Roughly speaking, a GNN iteratively maps a node to a feature vector via aggregating the feature vectors of the node's neighbors, and the last layer of the neural network outputs a label for the graph. Note that the graph does not need to be connected, i.e., GNN can still predict a label for a graph even if the graph consists of multiple disconnected components.

\myparatight{Training GNNs for graph classification} Suppose we are given a training dataset $\mathcal{D}_{tr}$=$\{(G_1, y_1), (G_2, y_2),\cdots, (G_{N}, y_{N}) \}$, where $G_i$ and $y_i$ respectively are the $i$th training graph and its true label, $i=1,2,\cdots,N$. 
Stochastic gradient descent (SGD) is often used to learn a GNN classifier from a training dataset. In particular, we randomly sample a batch of training graphs and compute the gradient of the loss function with respect to the model parameters for the batch. The loss function is usually cross-entropy loss. Then, we move the model parameters  towards the inverse direction of the gradient by a small step. Formally, we have $\theta = \theta - lr \cdot \nabla_{\theta}\mathcal{L}(\theta;batch)$, where $\theta$ is the model parameters of the GNN classifier $f$, $lr$ is called learning rate, $\nabla_{\theta}$ is gradient with respect to $\theta$, $\mathcal{L}$ is loss function, and $batch$ is a batch of training graphs randomly sampled from the training dataset.  This process is repeated until convergence or the maximum number of iterations is reached. The learnt GNN classifier $f$ is then used to predict labels for testing graphs. 

\subsection{Threat Model}
Our threat model is largely inspired by backdoor attacks in the image domain~\cite{gu2017badnets,chen2017targeted,liu2017trojaning,li2018hu,tran2018spectral,salem2020dynamic}. We characterize the threat model with respect to attacker's goal and attacker's capability.

\myparatight{Attacker's goal} An attacker has two goals. 
First, the backdoor attack should not influence the GNN classifier's accuracy on clean testing graphs, which makes the backdoor attack stealthy. If an attack sacrifices a GNN classifier's accuracy substantially, a defender could detect such low accuracy using a clean validation dataset and the GNN classifier may not be deployed. Second, the backdoored GNN classifier should predict an attacker-chosen target label for any testing graph once a trigger is injected to the testing graph.

\myparatight{Attacker's capability} We assume the attacker can poison some training graphs in the training dataset. Specifically, the attacker can inject a trigger to each poisoned training graph and change its label to an attacker-chosen target label. For instance, when the training graphs are crowdsourced from users, malicious users under an attacker's control can provide such poisoned training graphs.
Moreover,  the attacker can inject the same  trigger to testing graphs, e.g., the attacker's own testing graphs.
\section{Our Subgraph based Backdoor Attacks} 
\subsection{Attack Overview}
Figure~\ref{backdoorattack} illustrates the pipeline of our subgraph based backdoor attack. Our backdoor attack uses a  subgraph as a backdoor trigger. Suppose a subgraph consists of $t$ nodes. Injecting the subgraph to a graph means that we sample $t$ nodes from the graph uniformly at random, map them to the $t$ nodes in the subgraph randomly, and replace their connections as the subgraph. In the training phase, an attacker injects a subgraph/trigger to a subset of training graphs and changes their labels to an attacker-chosen target label. The training dataset with such injected triggers is called \emph{backdoored training dataset}. A GNN classifier is then learnt using the backdoored training dataset, and such GNN is called \emph{backdoored GNN}. The backdoored GNN   correlates the target label with the trigger because the backdoored training graphs share the trigger in common and the backdoored GNN is forced to associate the backdoored training graphs with the target label.  
In the testing phase, the attacker injects the same subgraph/trigger to a testing graph and the backdoored GNN  is very likely to predict the target label for the testing graph with trigger injected.

\begin{table*}[tp]\renewcommand{\arraystretch}{0.9} 
	
	\centering
	%\fontsize{6.5}{8}\selectfont
	\caption{Statistics of datasets.}
	\vspace{-3mm}
	\begin{tabular}{|c|c|c|c|c|c|c|c|c|c|c|c|c|}
		\hline
		\multirow{2}{*}{Datasets}&
		\multirow{2}{*}{\#Graphs}&
		\multirow{2}{*}{\#Classes}&
		\multirow{2}{*}{Avg. \#nodes}&
		\multirow{2}{*}{Avg. density}&
		\multicolumn{3}{c|}{\#Training graphs}&\multicolumn{3}{c|}{\#Testing graphs}\cr\cline{6-11}
		& & & & &Class 0&Class 1&Class 2&Class 0&Class 1&Class 2\cr
		\hline
		Bitcoin&658&2&11.53&0.342&219&219&-&110&110&-\cr\hline
		Twitter&1,481&2&63.10&0.523&489&498&-&245&249&-\cr\hline
        COLLAB&5,000&3&73.49&0.510&517&1,589&1,215&258&794&608\cr\hline
	\end{tabular}
	\label{three_dataset_statistics}
	\vspace{-4mm}
\end{table*}

\subsection{Attack Design}
\label{attackdesign}
Our backdoor attack involves injecting a backdoor trigger, i.e., a subgraph, to a graph. Designing the subgraph is key to our backdoor attack. Intuitively, the subgraph should be unique among the clean training/testing graphs, so the backdoored GNN is more likely to associate the target label with the subgraph. 
A naive method is to construct a complete subgraph (i.e., every pair of nodes in the subgraph is connected) as a backdoor trigger. However, such trigger could be easily detected especially when the number of nodes in the subgraph is large. For instance, a defender may search for complete subgraphs in a training or testing graph, and a complete subgraph may be detected as a backdoor trigger when complete subgraphs are unlikely to occur in the clean training/testing graphs.  Therefore, we propose to generate a random subgraph as backdoor trigger. 
In particular, we characterize our backdoor attack using four parameters: \emph{trigger size}, \emph{trigger density},  \emph{trigger synthesis method}, and \emph{poisoning intensity}. Next, we describe each of them. 
%Table~\ref{table:notation} shows some important notations used in our paper. 

\myparatight{Trigger size and trigger density} We call the number of nodes in the subgraph/trigger as trigger size. We denote the trigger size as $t$. 
%as the number of nodes in a graph that are trigger nodes. We use $t$ to denote it. 
 Given $t$ nodes,  there are $\frac{t \cdot ( t-1)}{2}$ pairs of nodes, which is the maximum number of edges that a subgraph with $t$ nodes could have. We define the trigger density of a subgraph as the ratio between the number of edges in the subgraph and the number of node pairs in the subgraph. We denote $\rho$ as the trigger density.  Formally, we have $\rho=\frac{2e}{t \cdot ( t-1)}$, where $e$ is the number of edges in the subgraph.

\myparatight{Trigger synthesis method} Given a trigger size $t$ and trigger density $\rho$, a trigger synthesis method generates a subgraph that has the given size and density. We generate a random subgraph using the Erdős-Rényi (ER) model~\cite{gilbert1959random}. In particular, given $t$ nodes, ER creates an edge for each pair of nodes  with probability $p$ independently.  $p$ is the expected density of the subgraph generated by ER. Therefore, we set $p = \rho$, which means that the generated subgraph has the given trigger density $\rho$ on average. 
In our experiments, we also evaluate triggers generated by the Small World (SW) model~\cite{watts1998collective} and Preferential Attachment (PA) model~\cite{barabasi1999emergence}, which are popular generative graph models developed by the network science community. Unlike ER, SW and PA generate subgraphs that are more similar to subgraphs in natural clean
graphs, e.g., they are small-world graphs and have power-law degree distributions. As a result, our backdoor attack with ER is more effective than that with SW and PA. 

In a nutshell, SW model first creates a ring in which each node is connected with its $k$ nearest neighbors. Then, for each edge in the ring, SW randomly rewires it with a certain probability, i.e., we move one of its end to a new node chosen uniformly at random from the rest of nodes with a certain probability. The parameter $k$ is related to the density of the subgraph. We set $k = \lceil(t-1) \rho\rceil$, with which the generated subgraph roughly has density $\rho$. PA adds nodes to the subgraph in a step-by-step manner. Initially, the subgraph has $k$ nodes and no edges. In each step, a new node is added to the subgraph and the new node is connected with randomly picked $k$ existing nodes in the subgraph, where the probability that a node is  picked is proportional to its degree. Intuitively, a new node prefers to connect with nodes who are already connected with many other nodes. The parameter $k$ is related to the density of the generated subgraph.  Formally, we set $k =\lceil \frac{t - \sqrt{t^2 - 2\cdot t \cdot (t-1) \cdot \rho}}{2} \rceil$, which allows the generated subgraph to roughly have density $\rho$. Note that PA  requires $\rho$ to be smaller than some threshold (i.e., the subgraph cannot be too dense) such that $k$ is a positive integer.

\myparatight{Poisoning intensity} Recall that our backdoor attack  poisons a subset of the training dataset by injecting the subgraph to some training graphs and changing their labels to the target label.  Poisoning intensity is the fraction of training graphs that are poisoned by the attacker. We denote by $\gamma$ the poisoning intensity.

\section{Attack Evaluation}

\subsection{Experimental Setup}
\label{exprimental_setup_attack}
\myparatight{Datasets} We evaluate our attacks on three publicly available real-world graph datasets.  Table~\ref{three_dataset_statistics} shows the statistics of our datasets. 

{\bf Bitcoin~\cite{weber2019anti}.} This dataset is used for graph-based fraudulent Bitcoin transaction detection. 
The original dataset has Bitcoin transactions collected at more than 40 different timestamps. Some transactions are manually labeled as illicit, some transactions are manually labeled as licit, while the remaining ones are unlabeled. We extracted $658$ labeled transactions. We represent each transaction as a graph. Specifically, in a graph, nodes are a transaction and the transactions that have Bitcoin flow with it and an edge between two transactions means that there was Bitcoin currency flow between them. Therefore, there are $658$ graphs and each graph has a label 0 or 1, which corresponds to illicit and licit transaction, respectively.

{\bf Twitter~\cite{wang2017sybilscar}.} This  dataset is used for graph-based fake user detection. In the original dataset, some users are labeled as fake, some are labeled as genuine, and the remaining are unlabeled.  We randomly picked $1,481$ labeled users. We represent each user using its ego network. In particular, in a user's ego network, the user and its followers/followees are nodes and an edge between two users indicates that they follow each other. A user's ego network is labeled as 0 if the user is fake and 1 otherwise. Therefore, this dataset includes  1,481 graphs and each graph has a label 0 or 1.

{\bf COLLAB \cite{yanardag2015deep}.}  COLLAB is a scientific collaboration dataset. A graph corresponds to a researcher's ego network, i.e., the researcher and its collaborators are nodes and an edge indicates collaboration between two researchers. A researcher's ego network  has three possible labels, i.e.,  High Energy Physics, Condensed Matter Physics, and Astro Physics, which are the fields that the researcher belongs to. The dataset has 5,000 graphs  and each graph has label 0, 1, or 2.

The Bitcoin and Twitter datasets represent GNN-based fraud detection. Both of them are binary classification tasks. We consider the COLLAB dataset because it is a widely used benchmark  for GNNs and it represents a multi-class classification task. For all three datasets, we extract a node's degree as its node feature. These diverse datasets can demonstrate the effectiveness of our backdoor attacks in different domains. 

\myparatight{Dataset splits and construction} We split each dataset to training dataset and testing dataset. Moreover, we construct backdoored training dataset and backdoored testing dataset via injecting a trigger to the graphs.    
In particular, we have the following datasets:

{\bf Clean training dataset.} For each dataset, we sample $2/3$ of the graphs uniformly at random as the training dataset. We call it \emph{clean training dataset}. 

 {\bf Clean testing dataset.} For each dataset, we treat the remaining graphs as \emph{clean testing dataset}.

 {\bf Backdoored training dataset.} Since our attack poisons some training graphs, we  construct a \emph{backdoored training dataset} from each clean training dataset. In particular, we randomly sample $\gamma$ fraction of graphs from a clean training dataset. Then, for each sampled training graph, we inject our backdoor trigger to it and relabel it as the target label. We assume label 1 as target label. In Bitcoin and Twitter, selecting label 1 as target label means evading fraud detection. 

 {\bf Backdoored testing dataset.} To evaluate the effectiveness of our attack, we create a backdoored testing dataset for each dataset. For each testing graph whose true label is not the target label, we inject our trigger to it. These testing graphs with injected trigger constitute our backdoored testing dataset.

\myparatight{GNN classifiers} Our attack does not rely on the architecture of GNN classifiers. We show our attacks for three popular GNN classifiers, i.e., GIN~\cite{xu2018powerful}, SAGPool~\cite{lee2019self}, and HGP-SL \cite{zhang2019hierarchical}. We use their publicly available implementations. When a classifier is learnt using a clean training dataset, we call the classifier \emph{clean classifier} and we denote it as $f_{c}$. When a classifier is learnt using a backdoored training dataset, we call the classifier \emph{backdoored classifier} and we denote it as $f_{b}$. Due to limited space, we show results on GIN unless otherwise mentioned.

\begin{figure*}[!t]
	 \vspace{-3mm}
	\centering
	{\includegraphics[width=0.29\textwidth]{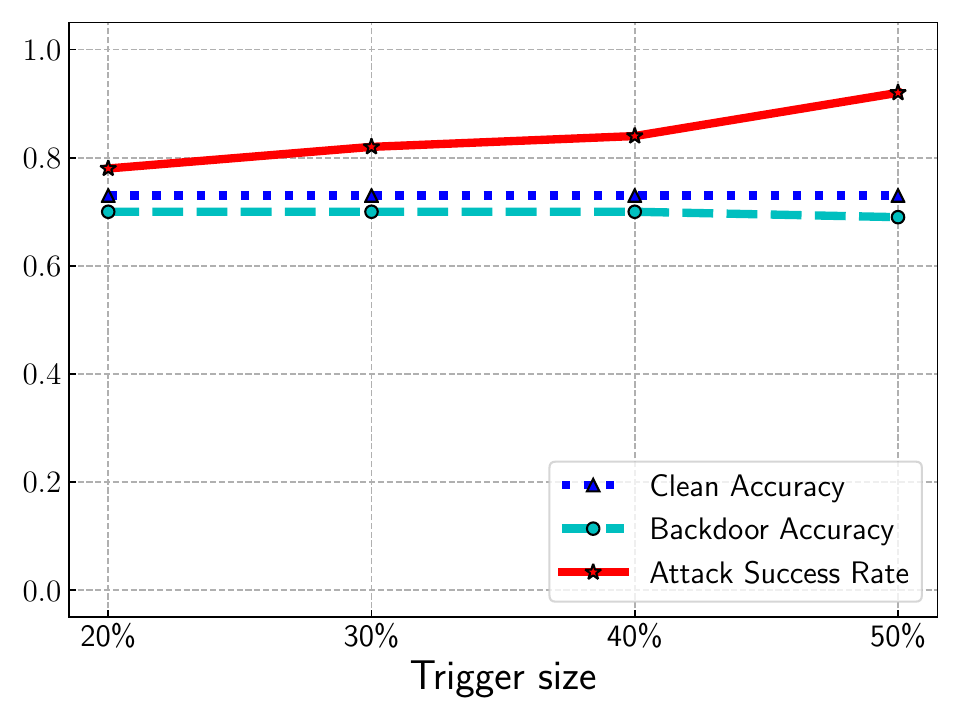}\label{impact_of_trigger_size_bitcoin}}
	{\includegraphics[width=0.29\textwidth]{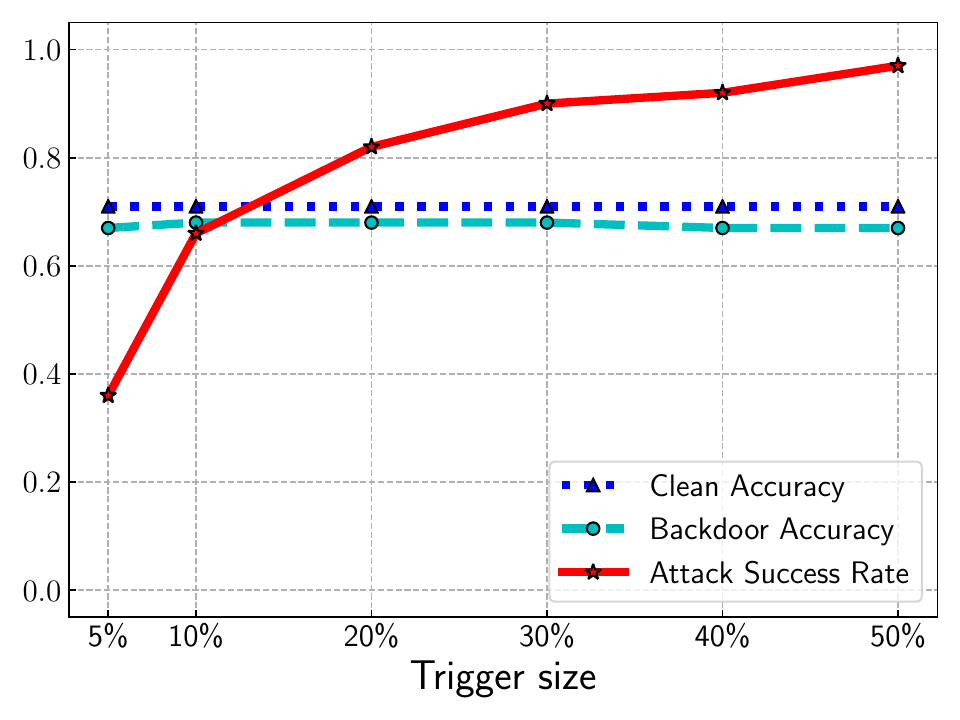}\label{impact_of_trigger_size_twitter}}
	{\includegraphics[width=0.29\textwidth]{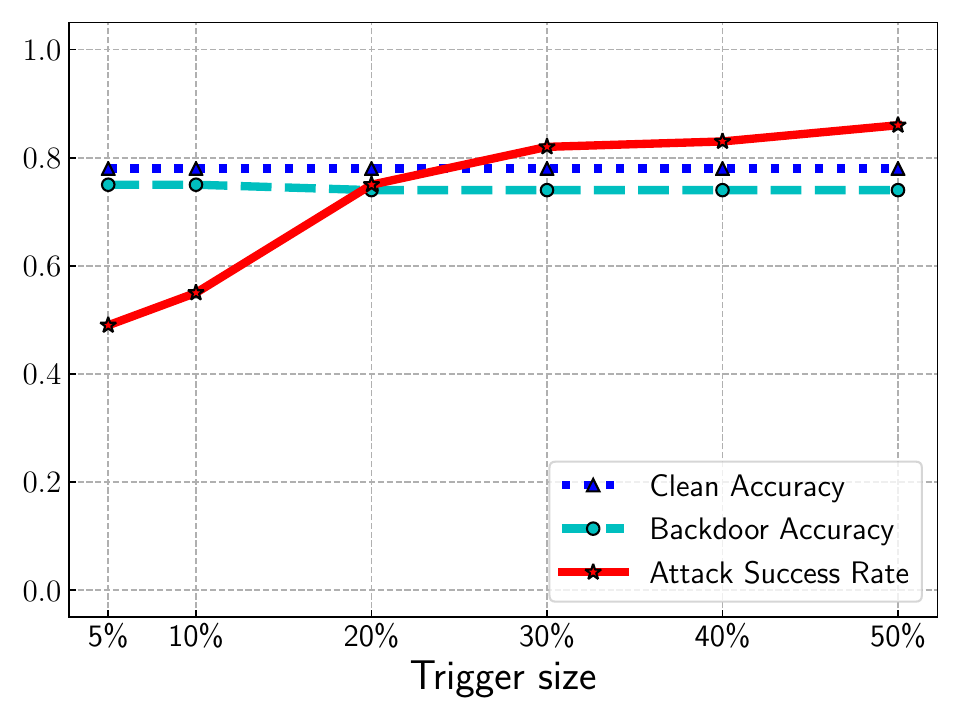}\label{impact_of_trigger_size_COLLAB}} \\
	{\includegraphics[width=0.29\textwidth]{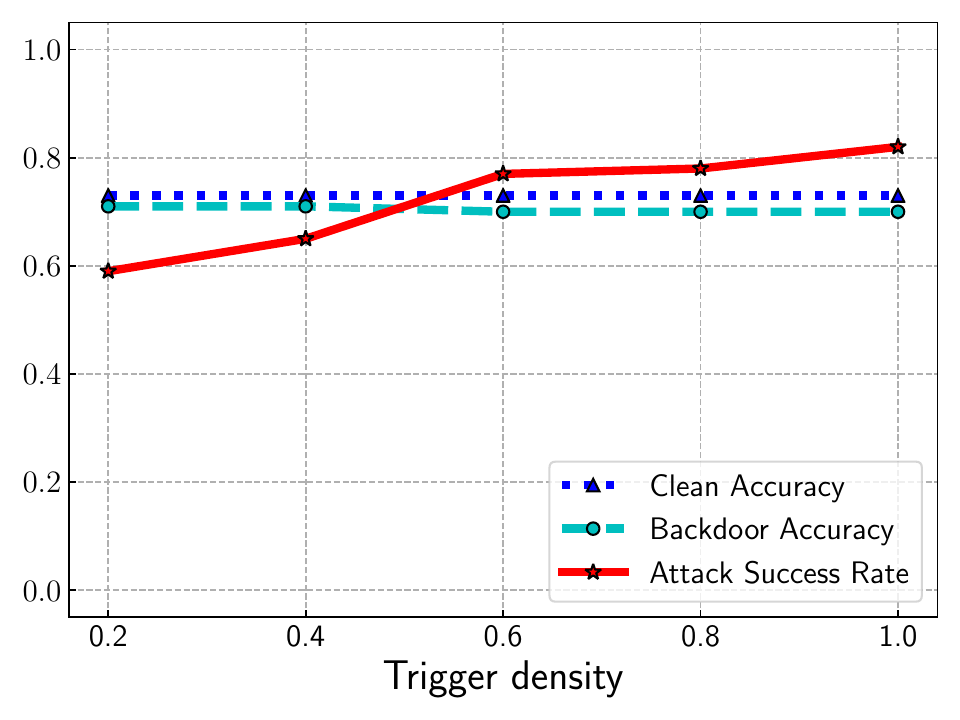}\label{impact_of_density_bitcoin}}
	{\includegraphics[width=0.29\textwidth]{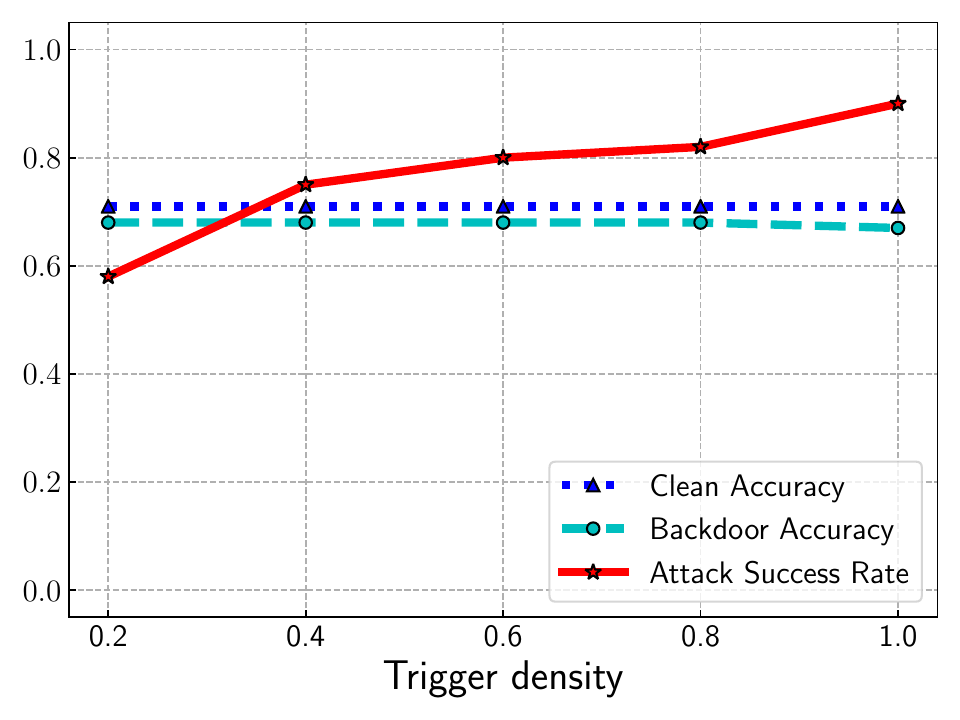}\label{impact_of_density_twitter}}
	{\includegraphics[width=0.29\textwidth]{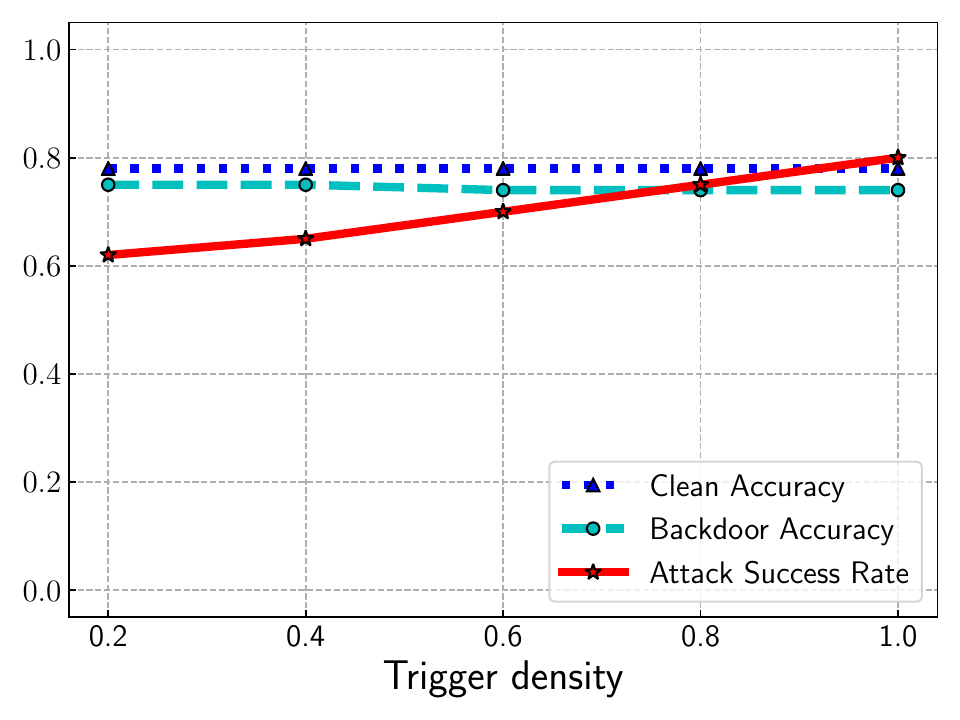}\label{impact_of_density_COLLAB}} \vspace{-3mm} \\
	\subfloat[Bitcoin]{\includegraphics[width=0.29\textwidth]{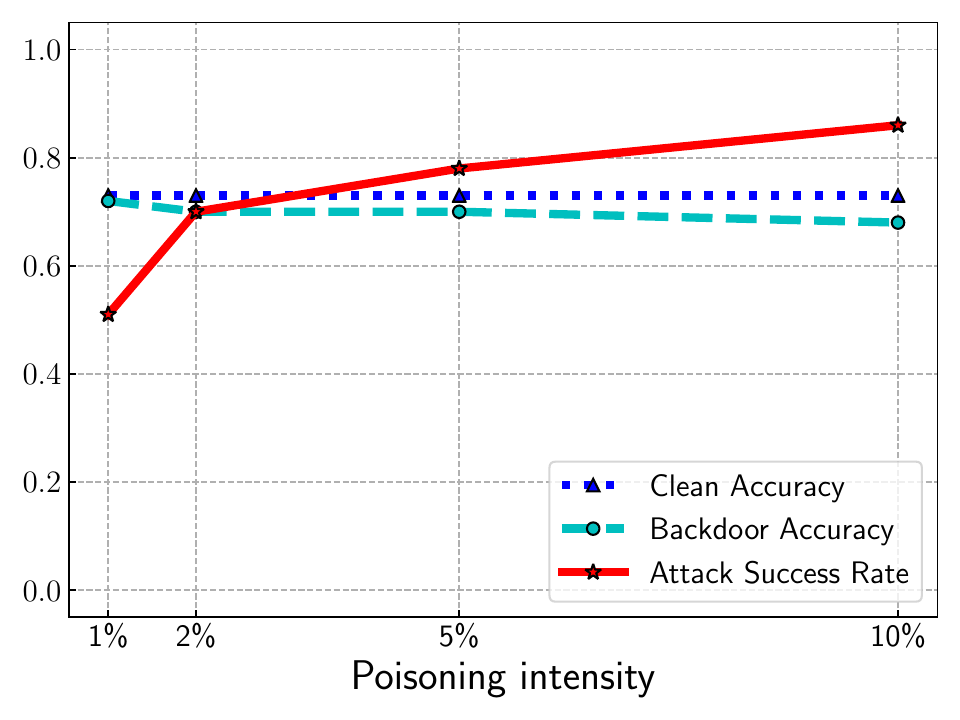}\label{impact_of_intensity_bitcoin}}
	\subfloat[Twitter]{\includegraphics[width=0.29\textwidth]{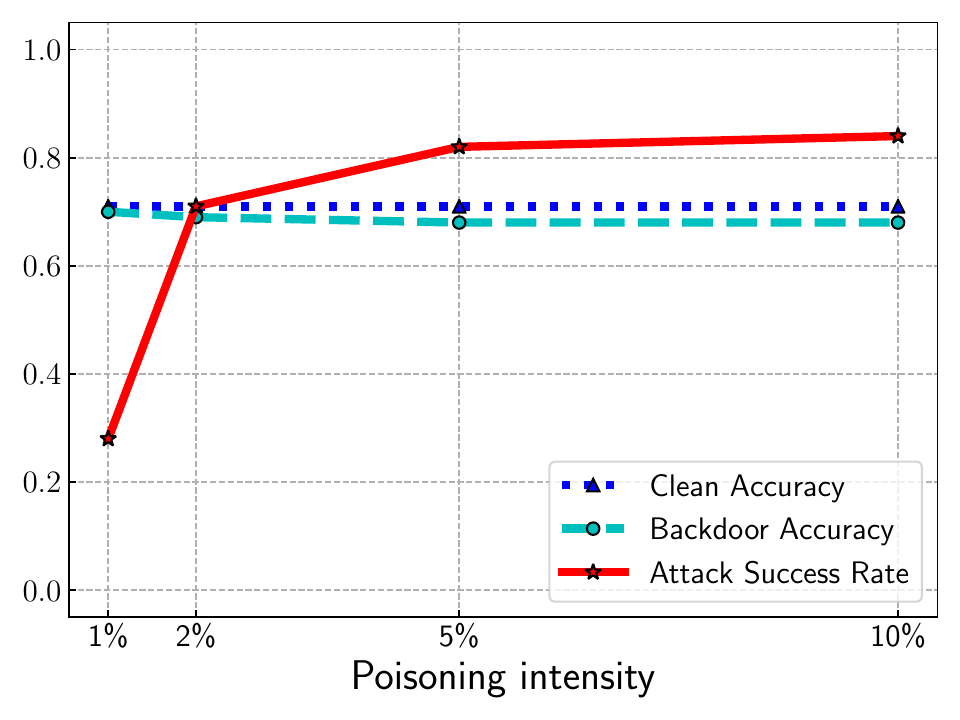}\label{impact_of_intensity_twitter}}
	\subfloat[COLLAB]{\includegraphics[width=0.29\textwidth]{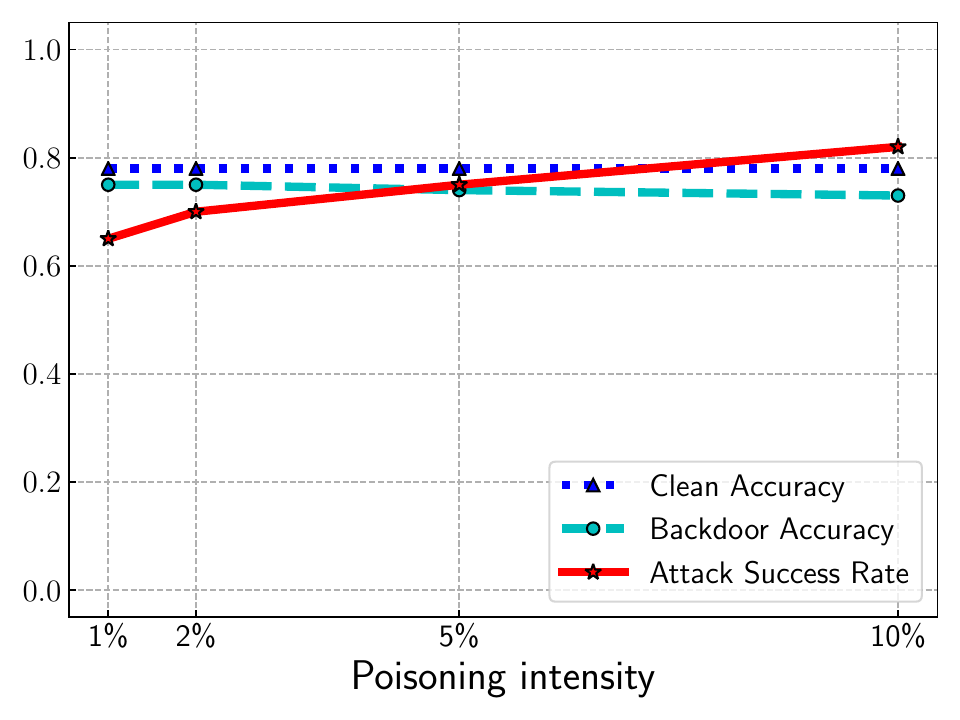}\label{impact_of_intensity_COLLAB}}
	\vspace{-3mm}
	\caption{Impact of trigger size (first row), trigger density (second row), and poisoning intensity (third row).}
	\label{impact_of_trigger_size_density_intensity}
	\vspace{-2mm}
\end{figure*}

\myparatight{Evaluation metrics} We use \emph{Clean Accuracy}, \emph{Backdoor Accuracy}, and \emph{Attack Success Rate} as evaluation metrics. Clean accuracy and backdoor accuracy respectively measure the accuracies of a clean classifier and a backdoored classifier for a clean testing dataset, while attack success rate is the fraction of graphs in the backdoored testing  dataset that are predicted to have the target label by a backdoored classifier. Next, we describe them in detail. %  the following three evaluation metrics: 

%\begin{itemize}
 {\bf Clean Accuracy.} Given a clean classifier $f_{c}$ and a clean testing dataset $\mathcal{D}_{ct}=\{(G_1,y_1),(G_2,y_2),\cdots,(G_n,y_n)\}$, we define the clean accuracy as the fraction of graphs in the clean testing dataset that are correctly predicted by the clean classifier $f_{c}$. Formally, we have the following: $\text{Clean Accuracy} = \frac{\sum_{i=1}^{n}\mathbb{I}(f_{c}(G_i)=y_i)}{n}$, where $\mathbb{I}$ is an indicator function.  
 
 {\bf Backdoor Accuracy.} The backdoor accuracy measures the accuracy of a backdoored classifier on the clean testing dataset. In particular, we define backdoor accuracy as the fraction of graphs in the clean testing dataset that can be correctly predicted by the backdoored classifier. Formally, we have: $\text{Backdoor Accuracy} = \frac{\sum_{i=1}^{n}\mathbb{I}(f_{b}(G_i)=y_i)}{n}$, where $\mathbb{I}$ is an indicator function. The difference between backdoor accuracy and clean accuracy measures the impact of our backdoor attack on accuracy for clean testing graphs. Recall that one of our attacker's goals is that the accuracy on clean testing graphs should not be influenced by our attack, i.e., backdoor accuracy and clean accuracy should be close.  

 {\bf Attack Success Rate.} 
Given a backdoored testing dataset $\mathcal{D}_{bt}=\{(G_1,y_1),(G_2,y_2),\cdots, (G_m,y_m)\}$, we define  attack success rate as the fraction of graphs in $\mathcal{D}_{bt}$ for which the backdoored classifier predicts the target label. Formally, we have the following:  $\text{Attack Success Rate} = \frac{\sum_{i=1}^{m}\mathbb{I}(f_{b}(G_i)=l)}{n}$, where $l$ is the attacker-chosen target label. 

\noindent {\bf Parameter setting:} Our attack has the following parameters: trigger size $t$, trigger density $\rho$, trigger synthesis method $M$, and poisoning intensity $\gamma$.  Different datasets have different graph sizes.  
Therefore, for each dataset, we set the trigger size $t$ to be $\varphi$ fraction of the average number of nodes per graph in the dataset (we use  ceiling  to obtain an integer number as the trigger size).  %Then, we can vary $\varphi$ to obtain different trigger sizes. 
Unless otherwise mentioned, we adopt the following default parameter settings:  $\varphi = 20\%$,  $\rho = 0.8$, $M=ER$, and $\gamma = 5\%$ in all three datasets. We will explore the impact of each parameter while fixing the remaining ones to their default settings. Note that when a graph has less nodes  than the trigger size, we replace the graph as the trigger.  ER may generate a subgraph/trigger with no edges as it randomly creates edges. When such case happens, we run ER multiple times until generating a subgraph with at least one edge. SW rewires an edge with a probability, which we set to be 0.8.

\begin{figure*}[!tbhp]
	% \vspace{-2mm}
	\centering
	\subfloat[Bitcoin]{\includegraphics[width=0.29\textwidth]{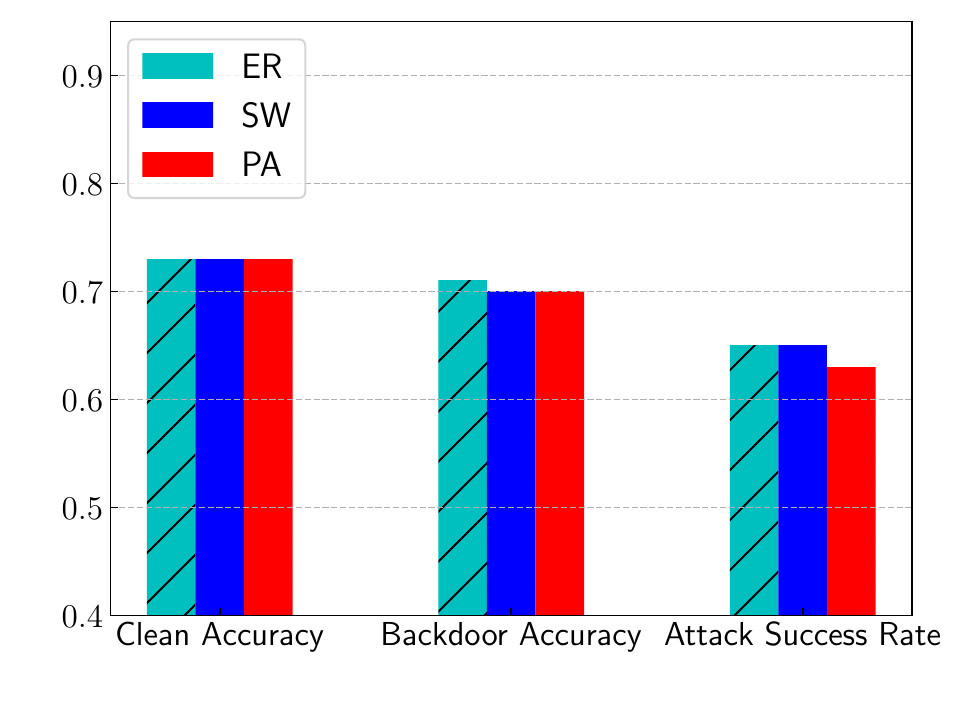}}
	\subfloat[Twitter]{\includegraphics[width=0.29\textwidth]{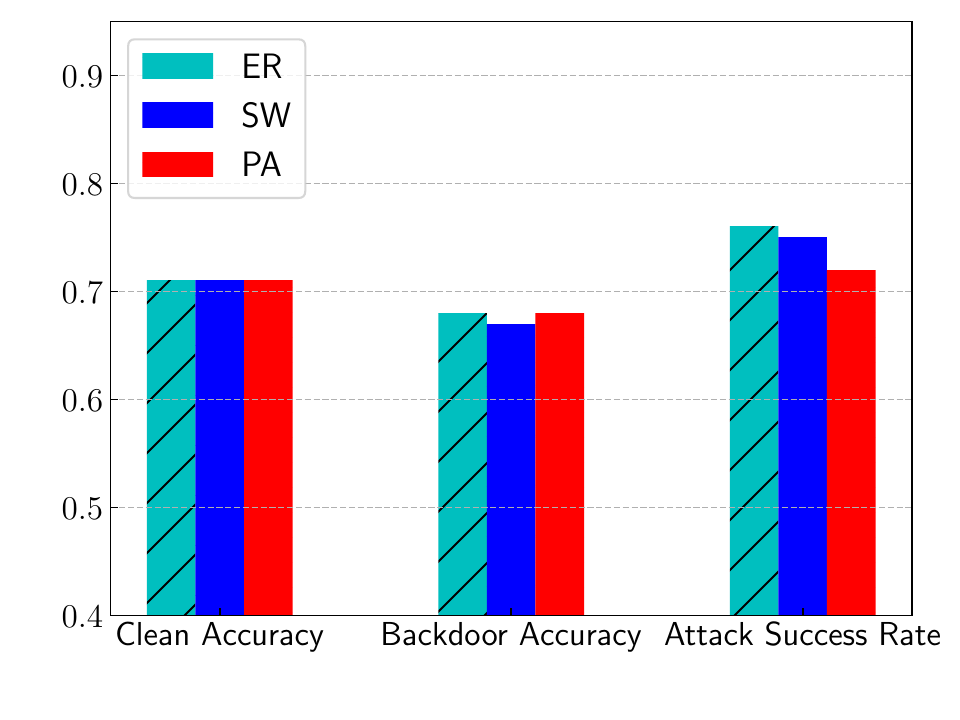}\label{impact_of_trigger_sysnthesis_method}}
	\subfloat[COLLAB]{\includegraphics[width=0.29\textwidth]{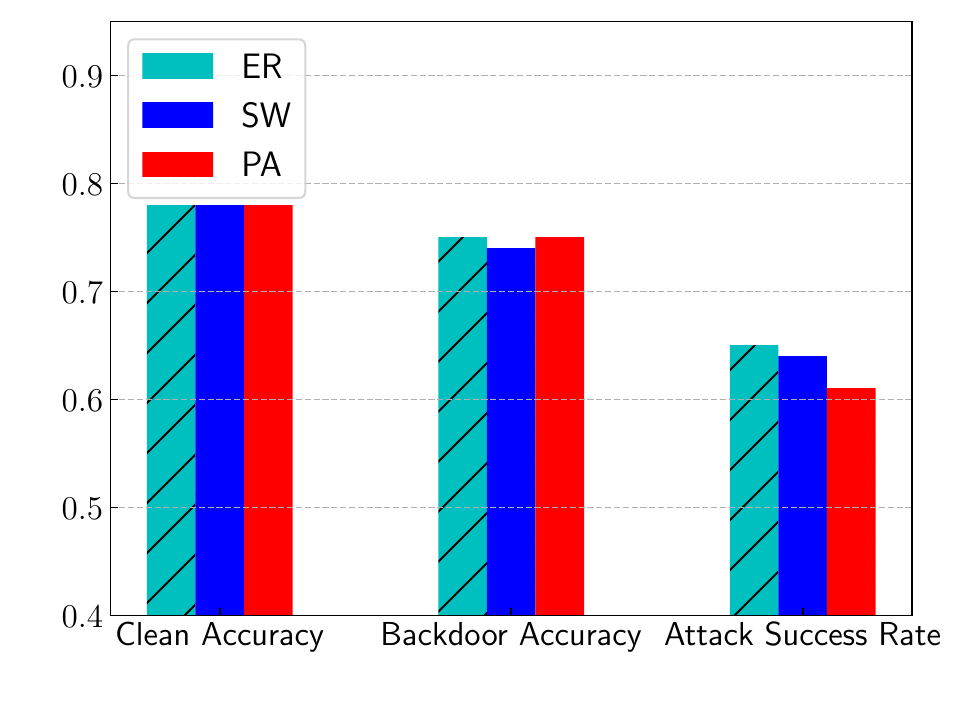}}
	\vspace{-3mm}
	\caption{Comparing trigger synthesis methods.}
	\vspace{-3mm}
	\label{impact_of_trigger_sysnthesis_method_other_two_datasets}
\end{figure*}

\begin{figure}[!t]
	% \vspace{-2mm}
	\centering
	\includegraphics[width=0.24\textwidth]{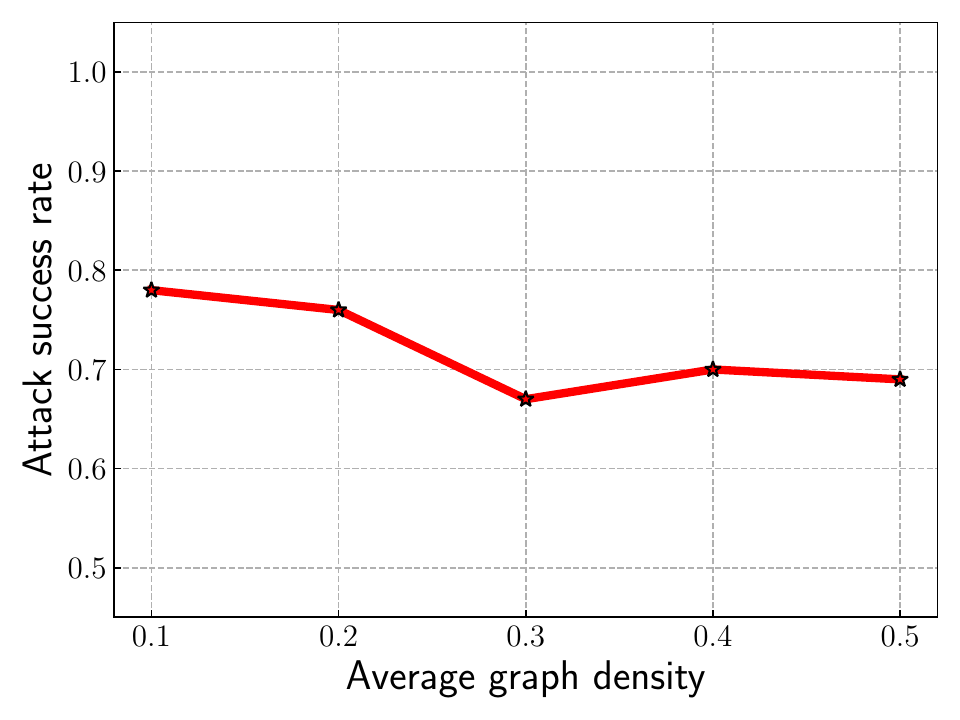}\label{datasetdensity}
	\vspace{-3mm}
	\caption{Attack success rate as a function of the average graph density on Twitter, where our trigger density is 0.3.}
	\vspace{-4mm}
	\label{success_rate_avg_graph_density}
\end{figure}

\subsection{Results}

\myparatight{Impact of trigger size, trigger density, and poisoning intensity}Figure~\ref{impact_of_trigger_size_density_intensity} shows
the impact of trigger size, trigger density, and poisoning intensity on the three datasets. First, we observe that our backdoor attacks have small impact on the accuracies for clean testing graphs. Specifically, backdoor accuracy is slightly smaller than clean accuracy. 
For instance, when the trigger size is 20\% of the average number of nodes per graph, the backdoor accuracy is 0.03 smaller than the clean accuracy on Twitter. Second, our backdoor attacks achieve high attack success rates and the attack success rates increase as the trigger size, trigger density, or poisoning intensity increases. The reason is that when the trigger size, trigger density, or poisoning intensity is larger, the backdoored GNN is more likely to associate the target label with the trigger. 

\myparatight{Comparing  trigger synthesis methods} 
Figure~\ref{impact_of_trigger_sysnthesis_method_other_two_datasets} compares ER, SW, and PA as trigger synthesis methods on the three datasets, where we set $\rho = 0.4$ since PA requires it to be small (see Section~\ref{attackdesign}). Our results show that ER  has higher attack success rates than SW and PA. We suspect the reason is that the subgraph generated by SW and PA is more similar to subgraphs in the clean graphs, e.g., they are small-world graphs and have power-law degree distributions, and thus the backdoored GNN is less likely to associate the target label with the subgraph. 

\myparatight{Impact of graph density} Intuitively, for a given trigger, the effectiveness of our backdoor attack may depend on 
the density of the clean training/testing graphs. To study the impact of the density of clean graphs, we randomly delete some edges in the Twitter graphs such that the average graph density ranges from 0.1 to 0.5. Figure~\ref{success_rate_avg_graph_density} shows the attack success rate of our attack as a function of the average graph density, where we set the trigger density to be 0.3 and the trigger size to be 20\% of the average number of nodes per graph. We observe a decreasing trend of attack success rate as the average graph density increases. One exception is that our attack success rate has a ``local minimum'' at the point where the average graph density is the trigger density. In other words, among the average graph densities that are around the trigger density, our attack is the least effective when  the average graph density is the same as the trigger density. The reason is that it is harder   for GNN to distinguish between the trigger and other subgraphs in the clean graphs when they have the same density, and thus it is harder for the backdoored GNN to associate the trigger with the target label.

\myparatight{Injecting  trigger in training vs. testing graphs}  Backdoor attacks inject a trigger to some training graphs and also testing graphs. One natural question is how successful a backdoor attack is if we only inject the trigger to the training graphs or testing graphs alone. Table~\ref{asr-train-test} shows the attack success rates of our backdoor attacks when injecting the trigger to only training graphs, only testing graphs, and both. We denote by $\mathcal{D}_{c}$ the set of clean testing graphs whose true labels are not the target label.   
Attack Success Rate-Baseline is the fraction of clean testing graphs in $\mathcal{D}_{c}$ that are predicted to have the target label by the clean GNN. Attack Success Rate-Baseline measures an attacker's success rate without injecting a trigger to any training/testing graph. Attack Success Rate-Train is the fraction of clean testing graphs in $\mathcal{D}_{c}$ that are predicted to have the target label by the backdoored GNN. Attack Success Rate-Test is the fraction of  testing graphs in $\mathcal{D}_{c}$ that are predicted to have the target label by the clean GNN when injecting the trigger to them. Attack Success Rate corresponds to our attack that injects the trigger to both training and testing graphs. 

We observe that injecting trigger to both training and testing graphs does improve attack success rates substantially. We also observe that injecting trigger to either training graphs or testing graphs alone increases the attack success rate upon the baseline.  This is because injecting trigger to training or testing graphs makes the GNN classifiers less accurate. For  Bitcoin and Twitter, being less accurate  is equivalent to higher  attack success rate since the two datasets are binary classification. For COLLAB,  the GNN classifiers are biased to be more likely to predict label 1 (i.e., target label) when making an incorrect prediction because label 1 has more training graphs (see Table~\ref{three_dataset_statistics}), and thus being less accurate increases the attack success rate. 

\begin{table}[tp]\renewcommand{\arraystretch}{1.0} 
	
	\centering
	%\fontsize{6.5}{8}\selectfont
	\caption{Attack success rates when injecting the trigger to only training graphs, only testing graphs, and both.}
	\vspace{-3mm}
	\begin{tabular}{|c|c|c|c|}
		\hline
		& Bitcoin & Twitter & COLLAB \\ \hline
				Attack Success Rate-Baseline &0.40  &0.19 &0.25  \\ \hline
		Attack Success Rate-Train &0.45  &0.28 &0.37  \\ \hline
		Attack Success Rate-Test &0.51  &0.24 &0.52  \\ \hline
		Attack Success Rate &0.78  &0.82 &0.75  \\ \hline
	\end{tabular}
	\label{asr-train-test}
	\vspace{-2mm}
\end{table}

\begin{table}[tp]\renewcommand{\arraystretch}{1.0} 
	
	\centering
	%\fontsize{6.5}{8}\selectfont
	\caption{Fixed trigger vs. random trigger on Twitter.}
	\vspace{-3mm}
	\begin{tabular}{|c|c|c|}
		\hline
		& Backdoor Accuracy & Attack Success Rate \\ \hline
		Fixed trigger &0.67  &0.82  \\ \hline
		Random trigger &0.66  &0.81  \\ \hline

	\end{tabular}
	\label{randomtrigger}
	\vspace{-3mm}
\end{table}

\myparatight{Fixed vs. random triggers} In all our experiments above, we use the same trigger in the training graphs and testing graphs. Table~\ref{randomtrigger} compares the backdoor accuracy and attack success rate when we use ER to generate one trigger and fix it (corresponding to ``Fixed trigger'') and when we use ER to generate a random trigger with the given trigger size and density for each poisoned training graph and testing graph (corresponding to ``Random trigger''). Our results show that  random trigger is nearly as effective as fixed trigger.  We suspect the reason is that the random triggers are structurally similar, e.g., they may be isomorphic, and a backdoored GNN can associate the structurally similar triggers with the target label.

\myparatight{Different GNN classifiers}
Table~\ref{GNNs} shows the attack results for three popular GNN classifiers on  Twitter. We observe that our attack is effective for different GNN classifiers. This is because  our attack does not rely on the architecture of GNN classifiers.

\begin{table}[]
	\centering
	%\fontsize{6.5}{8}\selectfont
	\caption{Our attack on different GNN classifiers on Twitter.}
	\vspace{-3mm}
	\begin{tabular}{|c|c|c|c|}
		\hline
		& GIN  & SAGPool & HGP-SL \\ \hline
		Clean Accuracy      & 0.71 & 0.69    & 0.72   \\ \hline
		Backdoor Accuracy   & 0.69 & 0.68    & 0.69   \\ \hline
		Attack Success Rate & 0.82 & 0.81    & 0.84   \\ \hline
	\end{tabular}
	\label{GNNs}
	\vspace{-3mm}
\end{table}

\myparatight{Comparing different ways to inject trigger} Our attack involves injecting a subgraph trigger to a training/testing graph. In particular, we pick $t$ nodes in a graph and replace their connections as the trigger, where $t$ is the trigger size.   One natural question is how to select the $t$ nodes in a graph. In all our above experiments, we  pick the $t$ nodes in a graph uniformly at random. We compare this \emph{random} strategy with three other strategies. Two strategies (called \emph{max degree} and \emph{min degree}) are to select the $t$ nodes with the largest and smallest degrees, respectively. The third strategy (called \emph{densely connected}) is to select  $t$ nodes that are densely connected, i.e., a set of $t$ nodes with the largest density, and we leverage the method in \cite{yuan2017exact} to find such $t$ nodes. Table~\ref{subgraph_selection} compares different strategies to select the $t$ nodes. We find that the random strategy has the most stable results. In particular, it achieves similar backdoor accuracy with other strategies on the three datasets. However, the random strategy achieves either much higher  attack success rates (e.g., on Twitter) or ones comparable with other strategies.

\section{Certified Defense}
\subsection{Overview}
Generally speaking, there are two types of defenses to build robust machine learning systems, i.e.,  \emph{empirical defenses} and \emph{certified defenses}. Empirical defenses are usually designed to defend against specific attacks and are often broken by strong adaptive attacks, which leads to a cat-and-mouse game between attackers and defenders. For instance, for backdoor attacks in the image domain, Salem et al.~\cite{salem2020dynamic} proposed dynamic backdoor attacks and showed it can bypass state-of-the-art empirical defenses~\cite{wang2019neural,gao2019strip,liu2019abs}. 
In Section~\ref{sec:discussion}, we show that an empirical defense based on dense subgraph detection is not effective for our attacks. Therefore, we focus on  certified defenses in this work. A certified defense provably predicts the same label for all data points in a certain region around an input. 

Randomized smoothing~\cite{cao2017mitigating,liu2018towards,lecuyer2018certified,li2019certified,cohen2019certified} is state-of-the-art technique to build provably robust machine learning. In particular, given an arbitrary classifier (called \emph{base classifier}), randomized smoothing builds a \emph{smoothed classifier} via randomizing an input, e.g., adding Gaussian noise to the input or randomly subsampling some features of the input. Roughly speaking, given an input, the smoothed classifier predicts the label that is the most likely to be returned by the base classifier when randomizing the input. Such predicted label for an input by the smoothed classifier certifiably remains the same when the $\ell_p$ norm of the perturbation added to the input is less than a certain threshold. 

Graph is essentially binary data, i.e., a pair of nodes can be either connected or unconnected. 
 For binary data, a randomized smoothing method called \emph{randomized subsampling}~\cite{levine2019robustness} achieves state-of-the-art certified robustness. Therefore, we design our certified defense  based on randomized subsampling. Next, we first introduce randomized subsampling and then discuss how to extend it to defend against our backdoor attacks. 

\begin{table}[!t]
	\centering
	%\fontsize{6.5}{8}\selectfont
	\caption{Comparing different ways to inject the trigger.}
	\vspace{-3mm}
	\setlength{\tabcolsep}{0.5mm}{
	\begin{tabular}{|c|c|c|c|c|}
		\hline
	{\small	Bitcoin      }       &{\small random} & {\small max degree }&{\small min degree }& {\small densely connected} \\ \hline
	{\small	Clean Accuracy }     & 0.73   & 0.73       & 0.73       & 0.73           \\ \hline
	{\small	Backdoor Accuracy }  & 0.7    & 0.71       & 0.71       & 0.69           \\ \hline
	{\small	Attack Success Rate} & 0.78   & 0.78       & 0.83       & 0.82           \\ \hline\hline
	{\small	Twitter   }          &{\small random }&{\small max degree }&{\small min degree} & {\small densely connected} \\ \hline
	{\small	Clean Accuracy}      & 0.71   & 0.71       & 0.71       & 0.71           \\ \hline
	{\small	Backdoor Accuracy }  & 0.69   & 0.7        & 0.69       & 0.7            \\ \hline
	{\small	Attack Success Rate} & 0.82   & 0.68       & 0.55       & 0.28           \\ \hline\hline
	{\small	COLLAB    }          &{\small random }&{\small max degree }&{\small min degree }& {\small densely connected} \\ \hline
	{\small	Clean Accuracy }     & 0.78   & 0.78       & 0.78       & 0.78           \\ \hline
	{\small	Backdoor Accuracy}   & 0.75   & 0.73       & 0.72       & 0.75           \\ \hline
	{\small	Attack Success Rate} & 0.76   & 0.76       & 0.74       & 0.76           \\ \hline
	\end{tabular}}
	\label{subgraph_selection}
	\vspace{-3mm}
\end{table}

\subsection{Randomized Subsampling}

%whose predictions are constant within a neighborhood of the input. 

\myparatight{Building a smoothed classifier via subsampling} Suppose we have a $s$-dimensional input $\mathbf{x}$ and a base classifier $h$ which maps $\mathbf{x}$ to a set of $c$ labels $\{1,2,\cdots,c\}$.  

Randomized subsampling creates a \emph{subsampled input} via keeping $z$ randomly subsampled features in $\mathbf{x}$ and setting the remaining features in $\mathbf{x}$ to a special value (e.g., 0). We denote such subsampled input as $\mathcal{S}(\mathbf{x},z)$. % to denote the randomly subsampled feature vector. 
Since  the subsampled input $\mathcal{S}(\mathbf{x},z)$ is random, the output of the base classifier $h$ for the subsampled input is also random. We denote $p_j$ as the probability that the base classifier $h$ outputs label $j$ when taking $\mathcal{S}(\mathbf{x},z)$ as input, i.e., $p_j = \text{Pr}(h(\mathcal{S}(\mathbf{x},z))=j), \forall j \in \{1,2,\cdots,c\}$. Then, randomized subsampling builds a smoothed classifier $\bar{h}$, which returns the label with the largest probability $p_j$ for the input $\mathbf{x}$. Formally, we have: 
\begin{align}
    \bar{h}(\mathbf{x}) = \argmax_{j\in \{1,2,\cdots,c\}} \text{Pr}(h(\mathcal{S}(\mathbf{x},z))=j)=\argmax_{j\in \{1,2,\cdots,c\}} p_j, 
\end{align}
where $\bar{h}(\mathbf{x})$ is the label that the smoothed classifier predicts for $\mathbf{x}$. In practice, to calculate the predicted label $\bar{h}(\mathbf{x})$, we create $d$ subsampled inputs from $\mathbf{x}$, use the base classifier to predict their labels, and take a majority vote among the $d$ labels as the predicted label $\bar{h}(\mathbf{x})$ (the majority vote label has the largest probability $p_j$).

\myparatight{Certified robustness} Suppose an attacker adds a perturbation $\delta$ to an input $\mathbf{x}$. The smoothed classifier certifiably predicts the same label for $\mathbf{x}$ when the $\ell_0$ norm of the perturbation is no larger than a threshold $R$, i.e.,  $\bar{h}(\mathbf{x} + \delta) = \bar{h}(\mathbf{x})$, $\forall ||\delta||_0 \leq R$. Moreover, $R$ is the maximum integer that satisfies the following inequality: 
\begin{align}
\label{certify_condition}
       {{s-R \choose z}} > (1.5 - \underline{p_l}){s \choose z}, 
\end{align}
where ${s-R \choose z}$ is the combination of $s-R$ things taken $z$ at a time, $l$ is the predicted label for $\mathbf{x}$ by the smoothed classifier (i.e., $l=\argmax_{j\in \{1,2,\cdots,c\}} p_j$), and $\underline{p_l}$ is a lower bound of $p_l$. 

\myparatight{Estimating $l$ and $\underline{p_l}$} To calculate $R$ in Equation~(\ref{certify_condition}), we need to know $l$ and $\underline{p_l}$, which can be estimated using a Monte-Carlo sampling method with a probabilistic guarantee~\cite{cohen2019certified}. Specifically, we randomly create $d$ subsampled inputs from $\mathbf{x}$. We use the base classifier to predict the labels of the $d$ subsampled inputs. The smoothed classifier takes a majority vote among the $d$ labels, i.e., the most frequent label among the $d$ subsampled inputs is predicted as the label $l$ for  $\mathbf{x}$. Moreover, we denote by $d_l$ the number of subsampled inputs for which the base classifier predicts label $l$. Theoretically, 

 $d_l$ follows a binomial distribution with parameters $d$ and $p_l$. Therefore, according to the Clopper-Pearson method~\cite{clopper1934use}, a lower bound $\underline{p_l}$ of  $p_l$ can be estimated with a confidence level $1-\alpha$ as follows: 
\begin{align}
\label{estimatepl}
    \underline{p_l} = B(\alpha; d_l, d- d_l +1), 
\end{align}
where $B(\alpha;\nu,\mu)$ is the $\alpha$th quantile of the Beta distribution with shape parameters $\nu$ and $\mu$.

\subsection{Defending against our Backdoor Attacks}
\label{defendourattack}
We leverage randomized subsampling to defend against our backdoor attacks. Next, we discuss how to predict label for a testing graph using a \emph{smoothed GNN} classifier, the certified robustness guarantee of the smoothed GNN classifier, and our method of training a base GNN classifier to improve the accuracy and robustness of the smoothed GNN classifier.

\myparatight{Smoothed GNN} Suppose we are given an GNN classifier (called \emph{base GNN classifier}) and a testing graph $G$. The base GNN classifier can be a backdoored GNN classifier. We can represent the structure of a graph as a binary vector (called \emph{structure vector}), where each entry of the vector corresponds to the connection status (connected or unconnected) of a pair of nodes in the graph. We view the structure vector as an input $\mathbf{x}$ in randomized subsampling. The smoothed GNN classifier predicts label for the testing graph $G$ following three steps. First, we create $d$ \emph{subsampled graphs} from the testing graph $G$. Specifically, to create a subsampled graph, we randomly sample $z$ entries in the testing graph's structure vector, keep their values, set the remaining entries of the structure vector to be 0, and convert the perturbed structure vector to a graph. Note that when a node feature (e.g., node degree) is derived from the graph structure, the feature should also be recalculated for nodes in the subsampled graph. 
We set $z$ as $\beta$ fraction of the entries in the testing graph's structure vector in our experiments, i.e., $z=\lceil \beta|G|(|G|-1)/2 \rceil$. We call $\beta$ \emph{subsampling ratio}.  Second, we use the base GNN classifier to predict labels of the $d$ subsampled graphs. Note that the base GNN can still predict a label even if a subsampled graph is disconnected. Third, the smoothed GNN classifier takes majority vote among the $d$ labels as the predicted label for the testing graph $G$. We denote the predicted label as $l$ and by $d_l$ the number of subsampled graphs that are predicted to have label $l$ by the base GNN classifier. 

\myparatight{Certified robustness} 
Intuitively, when the trigger is small, a majority of the $d$ subsampled graphs do not include edges in the trigger. Therefore, a majority of the predicted labels for the subsampled graphs are not influenced by the trigger and the majority vote among the predicted labels of the subsampled graphs (i.e., the label predicted by the smoothed GNN classifier) is not influenced by the trigger. 
Formally, the smoothed GNN classifier provably predicts the same label for a testing graph $G$ once the $\ell_0$ norm of the perturbation added to the graph's structure vector is bounded by $R$ in Equation~(\ref{certify_condition}), where $s=|G| (|G|-1)/2$ is the number of entries in the graph's structure vector. Our attack injects a trigger to a graph, which can be viewed as flipping some entries of the graph's binary structure vector. Therefore, the $\ell_0$ norm of the perturbation introduced by our attack is  the number of entries of a graph's structure vector that are flipped by our trigger, i.e., the number of edges that are deleted or added by our trigger. Injecting a trigger with size $t$ to a graph deletes or adds at most $t(t-1)/2$ edges in the graph. Therefore,  we know that the smoothed GNN classifier predicts the same label for a testing graph once the size of the trigger added to the testing graph is no larger than a threshold $T$. Formally, based on Equation~(\ref{certify_condition}) with $s=|G| (|G|-1)/2$ and $R=T(T-1)/2$,  we have  $T$ is the largest integer that satisfies the following:
\begin{align}
      {{|G| (|G|-1)/2 - T (T-1)/2 \choose z}} > (1.5 - \underline{p_l}){{|G| (|G|-1)/2 \choose z}},
\end{align}
where $\underline{p_l}$ is estimated using Equation (\ref{estimatepl}). We call the threshold $T$ \emph{certified trigger size}. When estimating $\underline{p_l}$, we set the confidence level $1-\alpha$ to be 0.999 in experiments.

\begin{algorithm}[!t]
    \caption{Training with Subsampling}
    \begin{algorithmic}[1]
    \REQUIRE $\mathcal{D}_{tr}$, $max\_iter$, $\beta$, and $lr$ (learning rate). \\
    \ENSURE $\text{Model parameter } \theta$ \\
    \STATE $\text{Initialize }\theta$ 
    \STATE  $iter = 0$ 
        \WHILE {$iter < max\_iter$} 
        \STATE $batch = \text{MiniBatch}(\mathcal{D}_{tr})$ 
        \STATE //we adapt the following two steps 
        \STATE ${\color{blue}subsampled\_batch = \text{Subsample\_batch}(batch,\beta)}$ 
        \STATE ${\color{blue}\theta = \theta - lr \cdot \nabla_{\theta}\mathcal{L}(\theta;subsampled\_batch)}$ 
        \STATE $iter = iter + 1$
        \ENDWHILE 
        \RETURN $\theta$ 
\end{algorithmic}
\label{algorithml1}
\end{algorithm}

\myparatight{Training base GNN classifier with subsampling} We can build a smoothed GNN classifier from any base GNN classifier, e.g., a backdoored GNN classifier. In our smoothed GNN classifier, the base GNN classifier is used to predict labels for subsampled graphs instead of the original testing graph. Therefore, the testing data distribution for the base GNN classifier is different from its training data distribution, which limits the accuracy of the base GNN classifier on the subsampled graphs and thus limits the accuracy and robustness of the smoothed GNN classifier. To overcome the distribution shift issue, we propose to train the base GNN classifier with subsampling.  
Algorithm~\ref{algorithml1} shows our training with subsampling. When using a random batch of training graphs to calculate the gradient of the loss function, we create a subsampled graph for each training graph in the batch and use the subsampled graphs to calculate the gradient and update the model parameters. Our experimental results demonstrate that training the base GNN classifier with subsampling significantly improves the accuracy and robustness of the smoothed GNN classifier. 

\begin{table}[!tbp]\renewcommand{\arraystretch}{1.0} 
	\setlength{\tabcolsep}{0.5pt}	
	\centering
	%\fontsize{6.5}{8}\selectfont
	\caption{Training with subsampling vs. training without subsampling on Twitter.}
	\vspace{-3mm}
	\begin{tabular}{|c|c|c|c|}
		\hline
		&{\small Clean  Accuracy} & {\small Backdoor Accuracy} & {\small Attack Success Rate} \cr\hline
		\makecell{\small Training without \\ subsampling} & 0.48 & 0.43 & 0.97 \cr\hline
		\makecell{\small Training with \\ subsampling} & 0.71 & 0.69 & 0.35 \cr\hline
	\end{tabular}
	\label{with_without_comparison}
	\vspace{-4mm}
\end{table}

\section{Defense Evaluation}

\subsection{Experimental Setup}
\myparatight{Datasets} We  also evaluate our defense on the three datasets, i.e., Bitcoin, Twitter, and COLLAB. Moreover, the dataset splits are the same as those in our attack evaluation in Section~\ref{exprimental_setup_attack}. 

\myparatight{Smoothed GIN classifiers} We consider GIN as the GNN classifier. When building our smoothed GIN classifier, we train the base GIN classifier with subsampling. Specifically, we train a \emph{clean base GIN classifier} and a \emph{backdoored base GIN classifier} using a clean training dataset and a backdoored training dataset, respectively. Then, we build a \emph{smoothed clean GIN classifier} and a \emph{smoothed backdoored GIN classifier} from them, respectively. We also train a \emph{clean GIN classifier} and a \emph{backdoored GIN classifier} using a clean training dataset and a backdoored training dataset, respectively. The clean/backdoored GIN classifiers are used as baselines to evaluate the performance of the smoothed clean/backdoored GIN classifiers. They are trained in the same way as those in our attack evaluation. In particular,  they are not trained with subsampling since they are not used to build smoothed classifiers.  

\myparatight{Evaluation metrics} We also consider Clean Accuracy, Backdoor Accuracy, and Attack Success Rate as our evaluation metrics. The clean accuracy of a smoothed clean GIN classifier  is the fraction of testing graphs in the clean testing dataset whose labels are correctly predicted by the classifier. The backdoor accuracy of a smoothed backdoored GIN classifier is the fraction of testing graphs in the clean testing dataset whose labels are correctly predicted by the classifier. The attack success rate of a smoothed backdoored GIN classifier is the fraction of  testing graphs in the backdoored testing dataset whose labels are  predicted as the target label by the classifier.

\begin{figure}[!tbhp]
	% \vspace{-2mm}
	\centering
\includegraphics[width=0.25\textwidth]{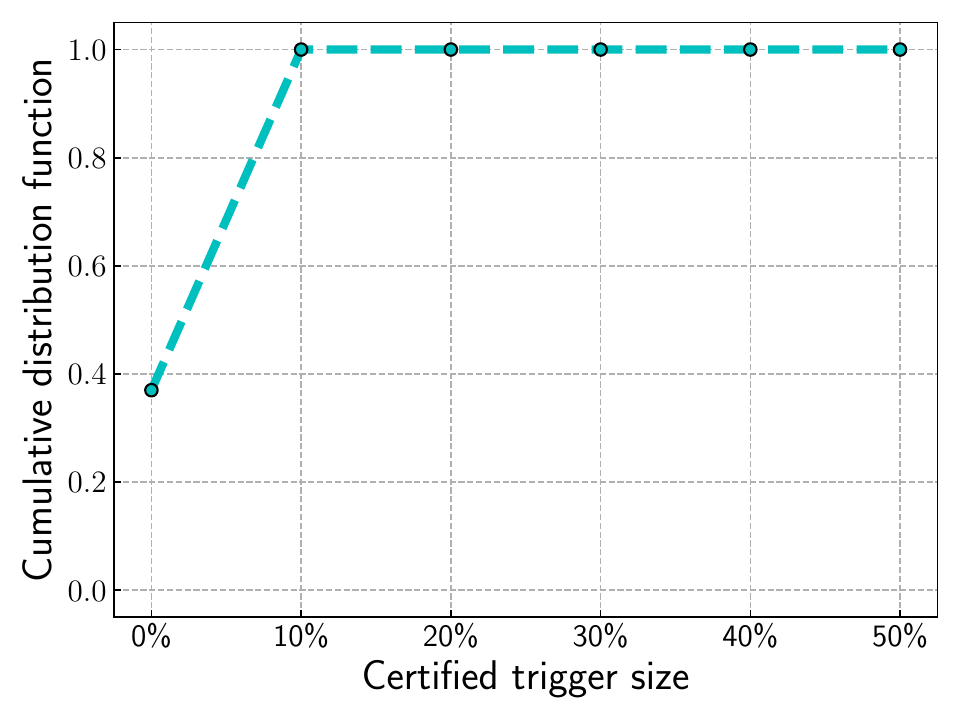}
\vspace{-3mm}
	\caption{Cumulative distribution function of the certified trigger size on Twitter dataset.}
	\label{different_cdf_three_datasets}
	\vspace{-6mm}
\end{figure}

\myparatight{Parameter setting} The attack parameter settings are the same as those in Section~\ref{exprimental_setup_attack}. Our defense has two parameters: number of subsampled graphs  $d$ and  subsampling ratio $\beta$. Unless otherwise mentioned, we adopt the following default setting: $d=100$ and $\beta = 10\%$ on all three datasets. We will explore the impact of one parameter while fixing the other parameter to its default setting.

\subsection{Results}
\label{defense-eva}

\myparatight{Training with subsampling vs. training without subsampling}\\ Table~\ref{with_without_comparison} shows the clean accuracy of our smoothed clean GIN classifier, and the backdoor accuracy and attack success rate of our smoothed backdoored GIN classifier, when the base GIN classifiers are trained with or without subsampling. We observe that training with subsampling substantially improves our smoothed classifiers. Moreover, the clean accuracy and backdoor accuracy are close, especially after training with subsampling. Therefore, we will only show backdoor accuracy in the remaining experiments for simplicity.

\begin{figure*}[!t]
\vspace{-2mm}
\centering
\subfloat[Bitcoin]{\includegraphics[width=0.29\textwidth]{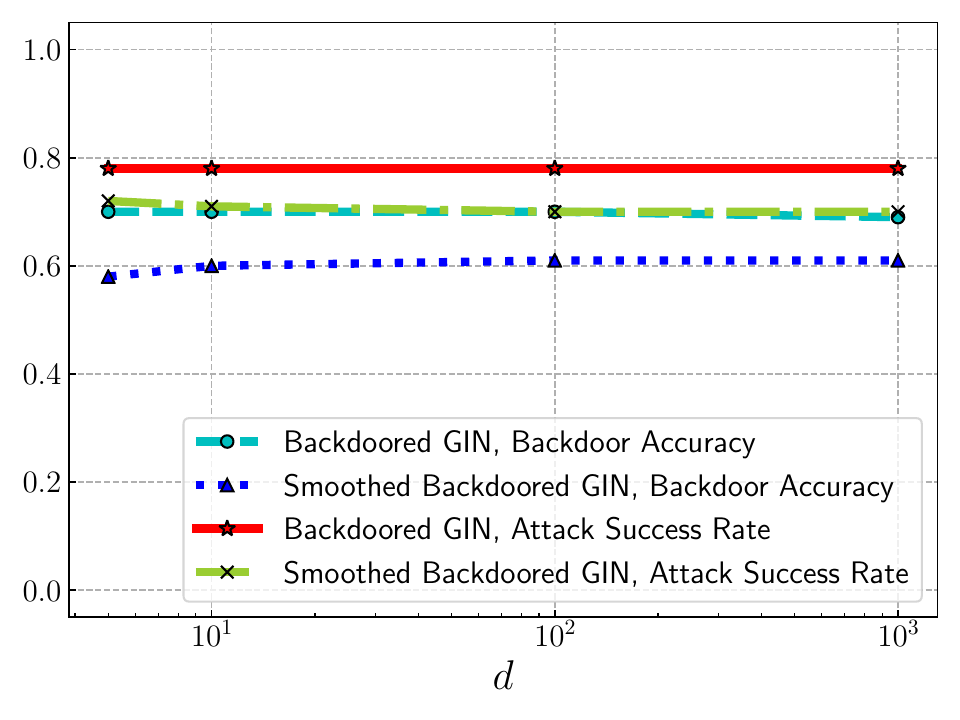}\label{different_d_bitcoin_three_datasets}}
\subfloat[Twitter]{\includegraphics[width=0.29\textwidth]{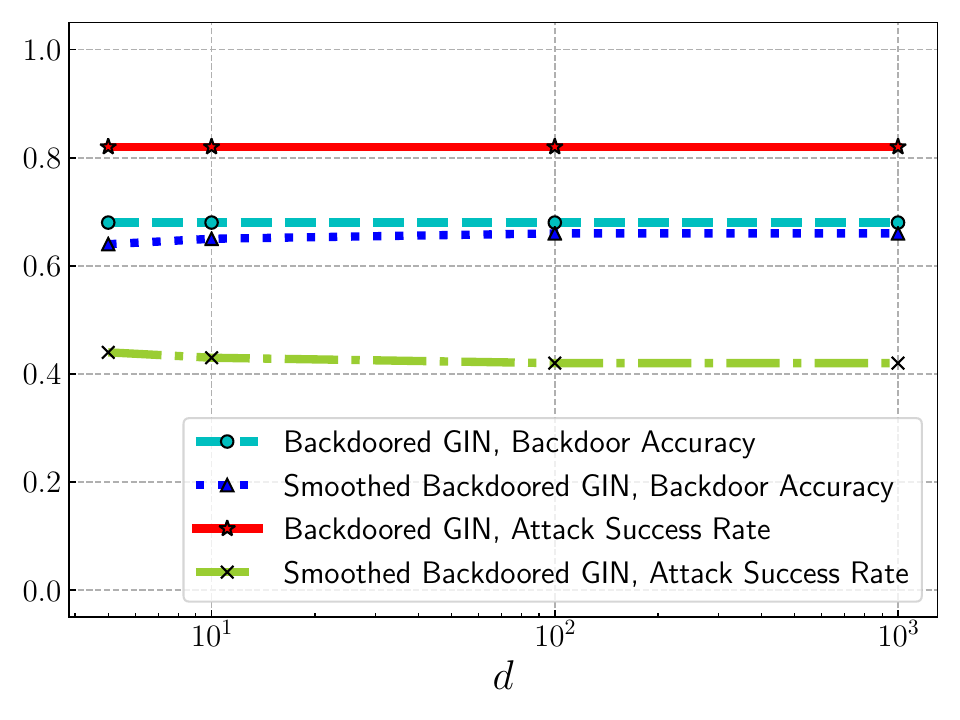}\label{different_d_twitter_three_datasets}}
\subfloat[COLLAB]{\includegraphics[width=0.29\textwidth]{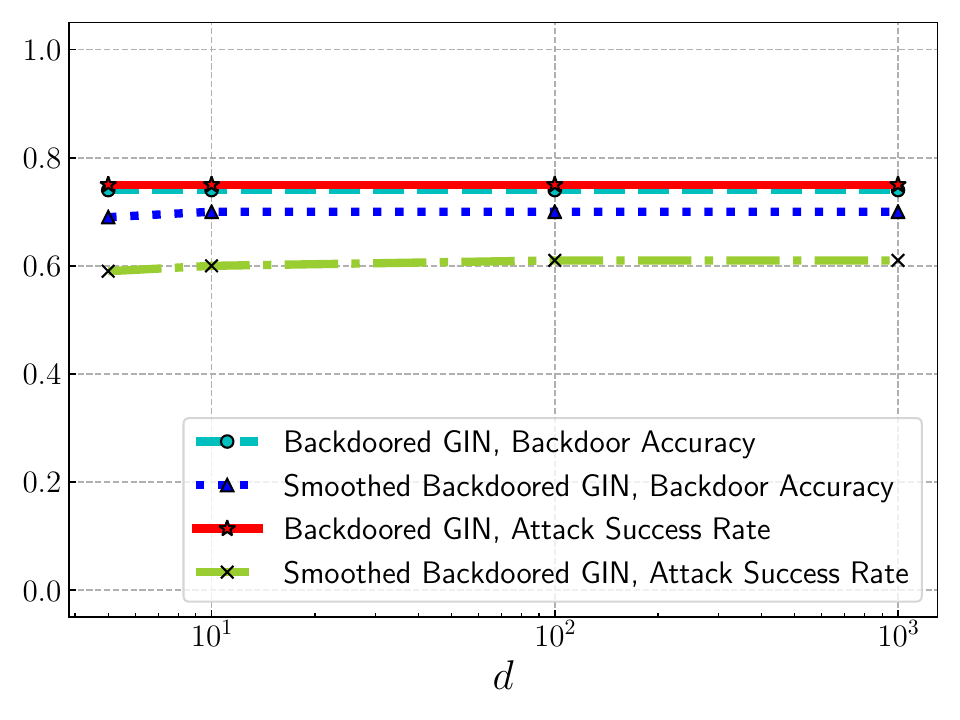}\label{different_d_collab_three_datasets}} \\ \vspace{-4mm}
\caption{Impact of the number of subsampled graphs $d$.}
\label{different_d_defense_three_datasets}
\vspace{-4mm}
\end{figure*}

\begin{figure*}[!t]
% \vspace{-2mm}
\centering
\subfloat[Bitcoin]{\includegraphics[width=0.29\textwidth]{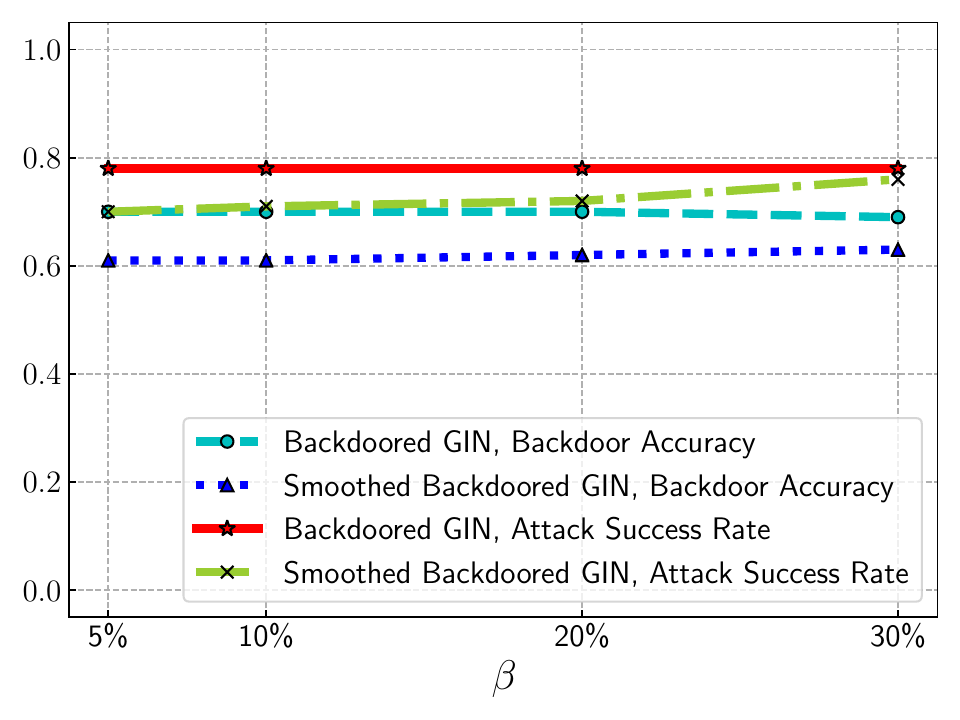}\label{different_beta_bitcoin_three_datasets}}
\subfloat[Twitter]{\includegraphics[width=0.29\textwidth]{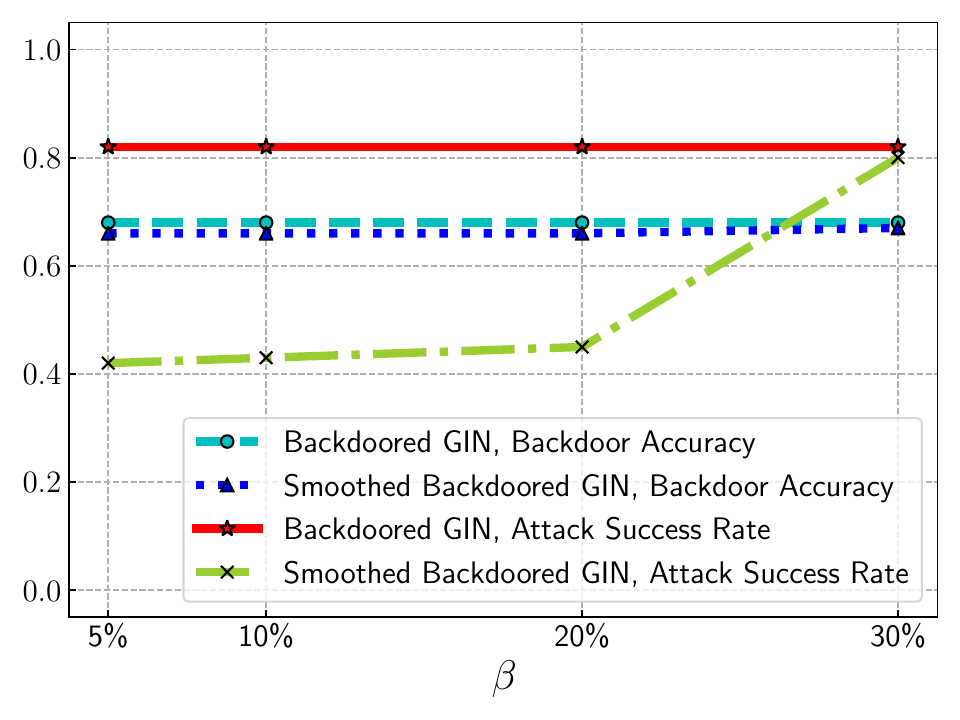}\label{different_beta_twitter_three_datasets}}
\subfloat[COLLAB]{\includegraphics[width=0.29\textwidth]{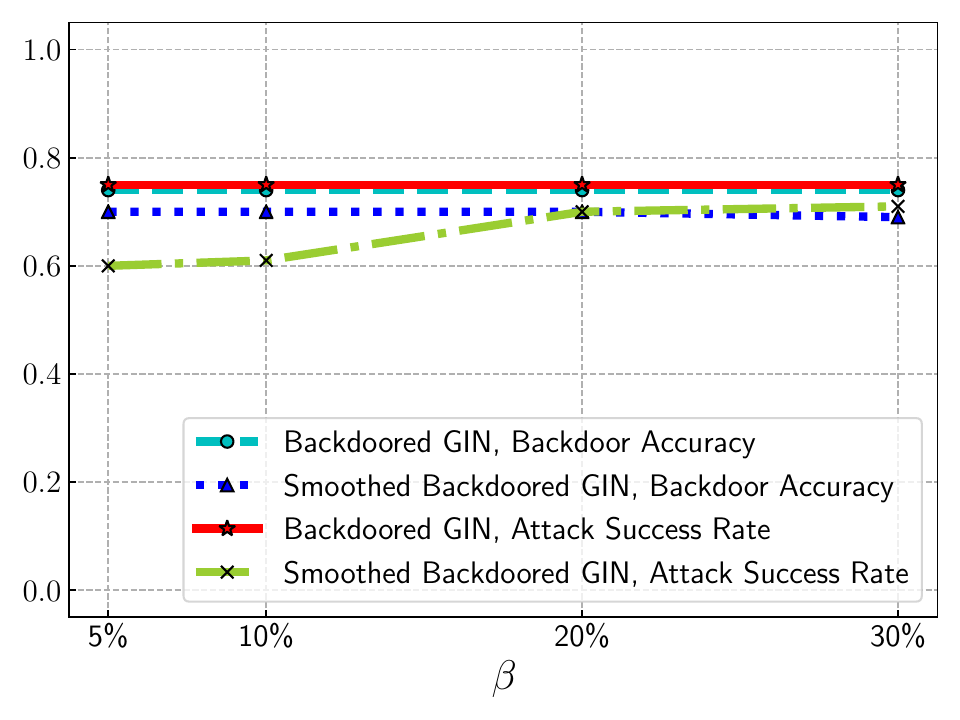}\label{different_beta_collab_three_datasets}} \\

\vspace{-4mm}
\caption{Impact of the subsampling ratio $\beta$.}
\label{different_beta_defense_three_datasets}
\vspace{-4mm}
\end{figure*}

\myparatight{Impact of the number of subsampled graphs $d$ and subsampling ratio $\beta$} Figure~\ref{different_d_defense_three_datasets} and Figure~\ref{different_beta_defense_three_datasets} show the impact of $d$ and $\beta$ on the backdoor accuracies and the attack success rates of the backdoored GIN classifier and our smoothed backdoored GIN classifier, respectively. The curves corresponding to  the backdoored GIN classifier are straight lines in the figures as they do not rely on $d$ nor $\beta$. 
Our results show that our smoothed classifiers achieve a tradeoff between backdoor accuracy and attack success rate. In particular, our smoothed backdoored GIN has lower backdoor accuracies than the backdoored GIN, but our smoothed backdoored GIN also has lower attack success rates. $d$ has a negligible impact on the smoothed backdoored GIN classifier when it is larger than 10. Our results indicate that our smoothed classifiers predict labels stably with only dozens of subsampled graphs. The reason may be that our datasets are binary or three-class classification problems.   As $\beta$ increases, our smoothed backdoored GIN classifier has higher backdoor accuracy but also higher attack success rate. The reason is that a higher subsampling ratio keeps more information of a testing graph in a subsampled graph but the subsampled graph is also more likely to include edges from the backdoor trigger. %We also have similar observations on the other two datasets, and thus we omit the results. 

\begin{figure*}[!t]
% \vspace{-2mm}
\centering
\subfloat[Bitcoin]{\includegraphics[width=0.29\textwidth]{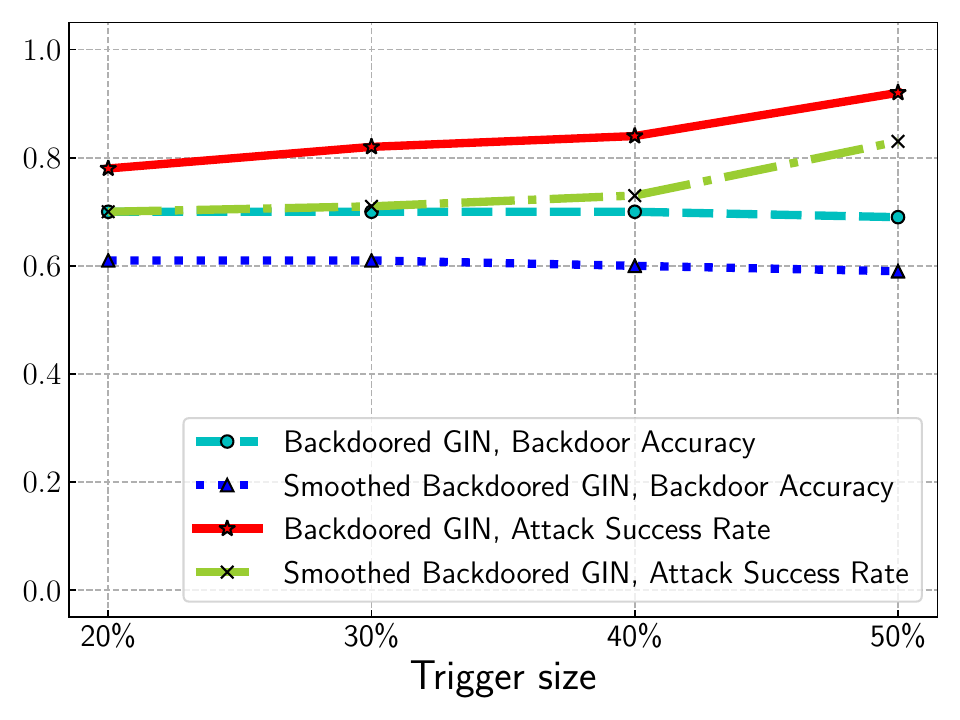}\label{different_trig_defense_bitcoin_three_datasets}}
\subfloat[Twitter]{\includegraphics[width=0.29\textwidth]{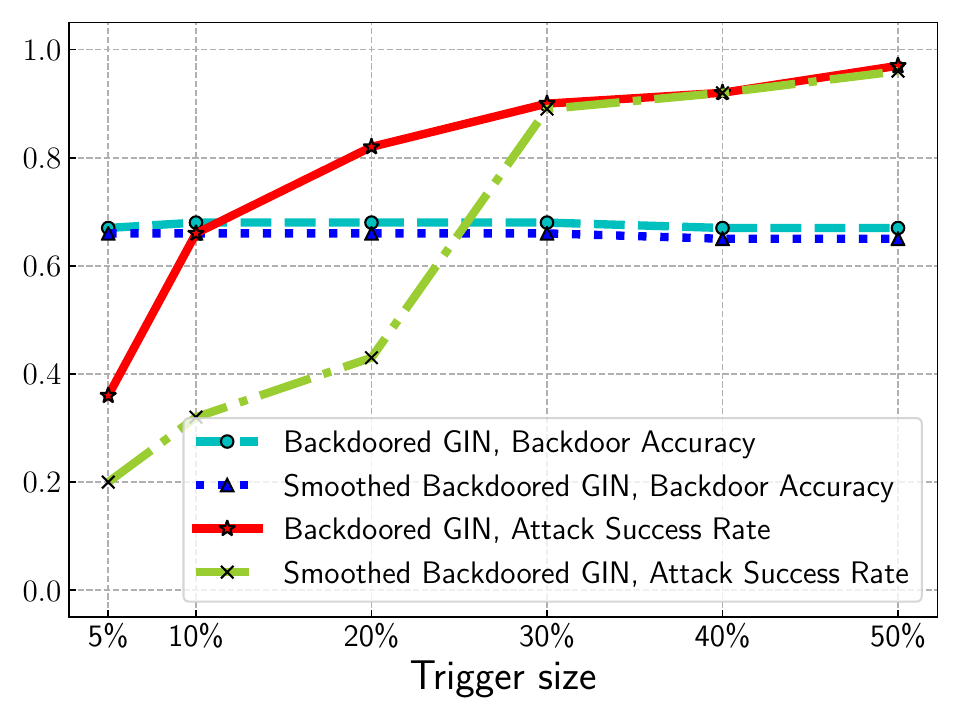}\label{different_trig_defense_twitter_three_datasets}}
\subfloat[COLLAB]{\includegraphics[width=0.29\textwidth]{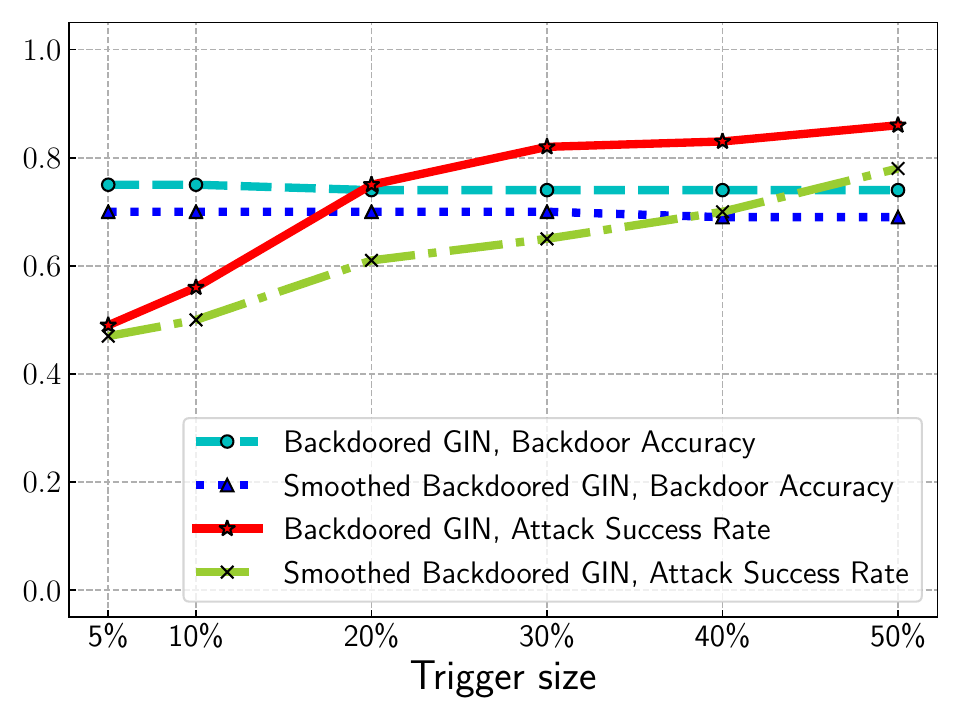}\label{different_trig_defense_collab_three_datasets}}
\vspace{-4mm}
\caption{Impact of the trigger size.}
\label{different_trig_defense_three_datasets}
\vspace{-3mm}
\end{figure*}

\myparatight{Impact of trigger size}
Figure~\ref{different_trig_defense_three_datasets} shows the impact of trigger size on the backdoor accuracy and attack success rate on the three datasets. When the trigger size is small, the smoothed backdoored GIN can reduce the attack success rate with some backdoor accuracy drop, compared to the backdoored GIN. In some scenarios, the attack success rate drops substantially with a small backdoor accuracy drop, indicating that randomized subsampling is an effective defense against our backdoor attacks. For instance, the smoothed backdoored GIN drops the attack success rate by 0.39 with only 0.02 backdoor accuracy drop on Twitter, when the trigger size is 20\% of the average number of nodes per graph. However, as the trigger size increases, the drops of the attack success rate become negligible in some scenarios. One notable example is that the smoothed backdoored GIN and backdoored GIN have almost the same attack success rate when the trigger size is 30\% of the average number of nodes per graph on Twitter. The reason is that the smoothed backdoored GIN has small certified trigger sizes. For instance, Figure~\ref{different_cdf_three_datasets} shows the cumulative distribution function of the certified trigger sizes of the testing graphs under the default parameter setting on Twitter. All testing graphs have trigger sizes less than $10\%$ of the average number of nodes per graph on Twitter. Our results highlight the needs of new defenses against our backdoor attacks, especially when the triggers are large.

\section{Discussion and Limitations}
\label{sec:discussion}

\myparatight{Detecting triggers via dense-subgraph detection} One potential way to defend against our backdoor attacks is to detect our trigger via dense-subgraph detection. This is an empirical defense and may be effective when our trigger is very dense, e.g., when our trigger is a complete subgraph. However, an attacker can use a sparser trigger to evade detection.  In particular, it may be hard to detect our trigger via dense-subgraph detection if our trigger is sparser than the clean graphs. However, our empirical results in  Figure~\ref{impact_of_trigger_size_density_intensity} (second row) and Figure~\ref{success_rate_avg_graph_density} show that our attacks are still effective even if our triggers are sparser than the clean graphs on average. Next, we further empirically evaluate detecting our trigger using dense-subgraph detection.

We adopt a state-of-the-art dense subgraph detection method \cite{hassen2017scalable} to detect our subgraph trigger. Specifically, given a graph with our trigger injected, we use the method to detect a dense subgraph in the graph. The method requires to specify the size of the dense subgraph. We assume the method knows our trigger size, which gives advantages to the detection method. Table~\ref{detection} shows the detection success rate, which is the fraction of training/testing graphs with our trigger injected whose detected dense subgraphs match our trigger. We observe that the detection success rate is very low on all three datasets.
When a dense subgraph is detected, the defender can remove its edges from the graph. Table~\ref{detection} also shows the backdoor accuracy and attack success rate when the defender removes the detected dense subgraph from each graph, where the same experimental settings in Section~\ref{exprimental_setup_attack} are used. We find that the backdoor accuracy drops significantly while our attack success rate remains high, demonstrating the ineffectiveness of the dense subgraph detection based defense.

\begin{table}[!tb]
	\centering
	%\fontsize{6.5}{8}\selectfont
	\caption{Dense subgraph detection based defense.}
	\vspace{-3mm}
	\begin{tabular}{|c|c|c|c|}
		\hline
		& Bitcoin & Twitter & COLLAB \\ \hline
		Detection Success Rate & 0.13    & 0.05     & 0.02    \\ \hline
		Backdoor Accuracy    & 0.53    & 0.58    & 0.59   \\ \hline
		Attack Success rate & 0.71    & 0.77    & 0.74   \\ \hline
	\end{tabular}
	\label{detection}
	\vspace{-4mm}
\end{table}

\myparatight{Applying our certified defense to image backdoor attacks} 
Our randomized subsampling based certified defense can be applied to image backdoor attacks.  
%Our method can be generalized to certify the backdoor attack on image domains. 
In particular, given a testing image and a base classifier (e.g., a backdoored neural network classifier), we can create multiple subsampled images from the testing image, use the base classifier to predict their labels, and take majority vote among the labels as the predicted label for the testing image. To create a subsampled image from the testing image, we randomly subsample some pixels, keep their values, and set the remaining pixels to a special value (e.g., 0). Such defense can certifiably predict the same label for a testing image when the size of the trigger injected to it is smaller than some threshold. We suspect that the defense may be effective in some scenarios, e.g., when the trigger is small. However, based on our empirical results in Section~\ref{defense-eva}, we suspect the defense may have limited effectiveness when the trigger is large.

\section{Related Work}

\myparatight{Backdoor attacks and their defenses in image domain}
Deep neural networks in the image domain were shown to be vulnerable to backdoor attacks~\cite{gu2017badnets,chen2017targeted,liu2017trojaning,li2018hu,clements2018hardware,tran2018spectral,yao2019latent,salem2020dynamic}. Specifically, a backdoored neural network classifier produces attacker-desired behaviors when a trigger is injected into a testing example. For instance, Gu et al.~\cite{gu2017badnets} proposed BadNets, which injects a backdoor trigger (e.g., a patch) to some training images and changes their labels to the target label. A neural network classifier trained on the backdoored training dataset predicts the target label for a testing image when the trigger is injected to it.  Liu et al.~\cite{liu2017trojaning} proposed to inject a backdoor to a neural network via fine tuning, which does not need to poison the training dataset.  Yao et al.~\cite{yao2019latent} developed latent backdoor attacks for transfer learning. 

To mitigate backdoor attacks, many defenses~\cite{chen2017targeted,liu2017trojaning,liu2017neural,liu2018fine,wang2019neural,gao2019strip,liu2019abs,guo2019tabor} have been proposed in the literature. Liu et al.~\cite{liu2018fine} proposed Fine-Pruning to remove  backdoor from a neural network via pruning its redundant neurons. Wang et al.~\cite{wang2019neural} proposed Neural Cleanse to detect and reverse engineer the trigger. Gao et al.~\cite{gao2019strip} tried to detect whether an input image includes a trigger or not via leveraging the input-agnostic characteristic of the backdoor trigger. Liu et al.~\cite{liu2019abs} proposed ABS to detect whether a neural network is backdoored or not via analyzing the behaviors of its internal neurons. We note that two  work~\cite{weber2020rab,wang2020certifying}, which are concurrent to ours, studied randomized smoothing based certified defenses against backdoor attacks in image domain. However, they use randomized smoothing with additive noise, e.g., Gaussian noise, uniform noise, or discrete noise, which has limited effectiveness at defending against backdoor attacks. For instance, Wang et al.~\cite{wang2020certifying} showed that the certified accuracy drops to $0$ when the attacker perturbs $3$ pixels on MNIST 1/7 dataset. We explored randomized subsampling based certified defense against our backdoor attacks to GNN. Our results show that such certified defense can reduce attack success rates with small accuracy drops when the trigger size is small, but it is less effective or ineffective when the trigger size is large.

\myparatight{Attacks to GNNs} Several studies~\cite{zugner2018adversarial,dai2018adversarial,bojchevski2019adversarial,wang2019attacking,zugner2019adversarial} showed that GNNs for node classification are vulnerable to adversarial structural perturbations. 
Specifically,  an attacker can perturb the graph structure such that a GNN based node classifier misclassifies many nodes in the graph indiscriminately or misclassifies some attacker-chosen nodes. For instance, Z{\"u}gner et al.~\cite{zugner2018adversarial} proposed an attack that can manipulate the graph structure while preserving important characteristics of the graph. Wang et al.~\cite{wang2019attacking} attacked collective classification via formulating the attack as an optimization problem and proposing several approximation techniques to solve the optimization problem. Moreover, their attacks can also transfer to GNN based node classifiers. Dai et al.~\cite{dai2018adversarial} proposed a reinforcement learning method to attack  GNNs for  both node and graph classification. For graph classification, their method perturbs a testing graph to be an  adversarial example such that a GNN misclassifies it. Chen et al.~\cite{ChenCCS17} proposed an attack for graph-based clustering. Our work is different from these studies because we focus on backdoor attacks to GNN based graph classification.

\myparatight{Randomized smoothing} Randomized smoothing~\cite{liu2018towards,cao2017mitigating,lecuyer2018certified,li2019certified,cohen2019certified,levine2019robustness,lee2019tight,jia2020certified} is state-of-the-art technique to build provably robust machine learning. Compared  with other certified defense mechanisms, randomized smoothing has two key advantages: 1) scalable to large neural networks, and 2) applicable to arbitrary classifiers. Randomized smoothing was initially proposed as an empirical defense~\cite{cao2017mitigating,liu2018towards}. For instance, Cao \& Gong~\cite{cao2017mitigating} proposed to use   uniform noise in a hypercube centered at a testing example to smooth the prediction for the testing example (they called their method \emph{Region-based Classification}). Lecuyer et al.~\cite{lecuyer2018certified} derived the first certified robustness guarantee for randomized smoothing with Gaussian or Laplacian noise via differential privacy. 
Cohen et al.~\cite{cohen2019certified} derived the first tight certified robustness guarantee for randomized smoothing with Gaussian noise by Neyman-Pearson Lemma~\cite{neyman1933ix}. Jia et al.~\cite{jia2020certified} generalized the tight certified robustness guarantee to general top-$k$ predictions for randomized smoothing with Gaussian noise.  
Jia et al.~\cite{jia2020certifiedcommunity} leveraged randomized smoothing to certify robustness of community detection against structural perturbations.  All these randomized smoothing methods add additive noise to a testing example. Levine et al.~\cite{levine2019robustness} proposed randomized subsampling, which does not use additive noise and achieves state-of-the-art $\ell_0$ norm certified robustness. We extend randomized subsampling to defend against our backdoor attacks. Our results show that randomized subsampling is ineffective in some scenarios.

\section{Conclusion and Future Work}
In this work, we showed that graph neural networks are vulnerable to backdoor attacks. 
Specifically, an attacker can inject a subgraph to some training graphs and change their labels to an attacker-chosen target label. A GNN classifier that is trained on the backdoored training dataset is very likely to predict the target label for any testing graph when the same subgraph is injected to it. Our empirical evaluation results on three real-world datasets show that our backdoor attacks achieve high success rates with a small impact on the GNN's accuracies for clean testing graphs. We also explored a randomized smoothing based certified defense against our backdoor attacks. Our empirical results show that the certified defense  is ineffective in some scenarios, highlighting the needs of new defenses against our backdoor attacks. Interesting future work includes: 1) detecting whether a GNN classifier is backdoored or not, and 2) designing new  defenses  against our backdoor attacks. 

\noindent\textbf{ACKNOWLEDGMENTS}\\
We thank the anonymous reviewers for insightful reviews. This work was supported by NSF grant No. 1937787.

{
\balance{
\bibliographystyle{ACM-Reference-Format}
\bibliography{refs}
}}

\end{document}